\documentclass{cleaniwsspa}
\usepackage{amssymb}
\usepackage[pdftex]{graphicx}
\usepackage{epstopdf}
\usepackage[dvipsnames]{xcolor}
\usepackage{amsmath}
\usepackage{hyperref}
\usepackage[utf8]{inputenc}
\usepackage{hyperref}
\usepackage[bottom]{footmisc} %
\usepackage{caption} %
\usepackage{booktabs} %
\usepackage{multirow} %
\usepackage{ulem} %
\usepackage{comment}
\usepackage{dblfloatfix}

\begin{document}

\markboth
  {Sound and Music Biases in Deep AMT Models} %
  {Mart\'ak et al.}

\title{Sound and Music Biases in Deep Music Transcription Models: \\
A Systematic Analysis}

\author{Luk\'a\v{s} Samuel Mart\'ak$^{*}$}{lukas.martak@jku.at}{1}
\author{Patricia Hu}{patricia.hu@jku.at}{1}
\author{Gerhard Widmer}{gerhard.widmer@jku.at}{1}

\affiliation{1}{Institute of Computational Perception \& LIT AI Lab}{Johannes Kepler University, Linz, Austria}

\def\thefootnote{*}\footnotetext{Corresponding author.}
\def\thefootnote{\arabic{footnote}}

\begin{abstract}
Automatic Music Transcription (AMT) --- the task of converting music audio into note representations --- has seen rapid progress, driven largely by deep learning systems.
Due to the limited availability of richly annotated music datasets, much of the progress in AMT has been concentrated on classical piano music, and even a few very specific datasets.
Whether these systems can generalize effectively to other musical contexts remains an open question.
Complementing recent studies on distribution shifts in sound (e.g., recording conditions), in this work we investigate the musical dimension --- specifically, variations in genre, dynamics, and polyphony levels. 
To this end, we introduce the MDS corpus, comprising three distinct subsets --- (1) Genre, (2) Random, and (3) MAEtest --- to emulate different axes of distribution shift.
We evaluate the performance of several state-of-the-art AMT systems on the MDS corpus using both traditional information-retrieval and musically-informed performance metrics.
Our extensive evaluation isolates and exposes varying degrees of performance degradation under specific distribution shifts.
In particular, we measure a note-level F1 performance drop of 20 percentage points due to sound, and 14 due to genre.
Generally, we find that dynamics estimation proves more vulnerable to musical variation than onset prediction.
Musically informed evaluation metrics, particularly those capturing harmonic structure, help identify potential contributing factors.
Furthermore, experiments with randomly generated, non-musical sequences reveal clear limitations in system performance under extreme musical distribution shifts.
Altogether, these findings offer new evidence of the persistent impact of the Corpus Bias problem in deep AMT systems.

\keywords automatic music transcription, musical distribution shift, corpus bias, robustness, evaluation benchmark, out-of-distribution inference, generalization, polyphonic piano transcription.
\end{abstract}

\section{Introduction}\label{sec:intro}

Automatic Music Transcription (AMT) is the task of converting audio signals into symbolic representations of music. It encompasses several variants, including multi-pitch estimation, frame-level and note-level transcription, and score-level transcription. The latter -- referred to as Complete Music Transcription \cite{ycart2020investigating} -- aims to produce fully notated scores, including quantized rhythms, pitch spelling, and time signatures, but remains relatively underexplored. In this article, we focus on note-level AMT, where the goal is to extract a MIDI-like representation describing pitch, onset, offset, and velocity. Specifically, we restrict our scope to polyphonic piano music.

A key motivation for this work is to gain a deeper understanding of the performance of state-of-the-art automatic music transcription (AMT) systems, particularly their inference performance under out-of-distribution (OOD) conditions. While many AMT models demonstrate impressive results on in-distribution (ID) data -- that is, the kind of data on which they were trained (which in many cases come from the MAESTRO dataset \cite{hawthorne2019enabling})
--, their robustness to data distribution shifts (such as when transcribing music outside the training domain) remains underexplored. This work investigates how AMT performance changes as the data distribution shifts from ID (e.g., MAESTRO) to OOD datasets which may differ in various aspects, including instrument timbre, recording conditions, and musical style. Importantly, we consider OOD as a multi-dimensional concept, with separable axes of distribution shift such as \textit{sound} (e.g., different instruments or recording environments) and \textit{music} (e.g., genre or tonal structure). By systematically analyzing these dimensions, we aim to provide a more nuanced understanding of AMT robustness and identify specific challenges that arise in real-world scenarios where the data may not match the training distribution.

In addition to standard quantitative recall/precision metrics as generally used in the AMT literature, we will make use of a new set of \textit{musically informed metrics} that have recently been introduced \cite{hu2024musically}, which provide complementary insights by capturing various musically and perceptually relevant facets of transcription quality that traditional evaluation metrics may overlook. These metrics, to be briefly described in Section \ref{sec:metrics_mi} specifically address aspects of transcription quality related to music performance and our perception of it, such as timing nuances, articulation, or dynamic accuracy. In this way, we hope that practitioners can gain a more nuanced and actionable understanding of how AMT systems perform across a broader spectrum of musically relevant dimensions.

We also compile and publish a new, compact benchmark dataset for the evaluation of automatic music transcription (AMT) systems.
It is designed with two key characteristics that set it apart from widely used benchmark datasets such as MAESTRO. First, it features a constant sound distribution, but from a real instrument: a (our) Yamaha Disklavier grand piano; this presents an \textit{out-of-distribution (OOD) sound profile} relative to MAESTRO, the dataset on which most state-of-the-art models are trained. Second, it offers a \textit{carefully controlled music distribution}, encompassing both a diverse range of musical genres and random note sequences, in contrast to the exclusively classical repertoire found in MAESTRO. This dual focus on sound and music distribution should provide a meaningful and challenging testbed for assessing the robustness and generalization capabilities of modern AMT systems.

Based on all this, we conduct a comprehensive quantitative and comparative analysis of several state-of-the-art AMT systems. This should enable practitioners and developers
to obtain deeper insights into these systems' behavior and limitations
under conditions that differ from the typical in-distribution (ID) sound and music data.
To further support reproducibility and progress in the field, we provide, in addition to the evaluation dataset itself\footnote{\url{https://zenodo.org/records/17467279}}, the source code used to generate the dataset and conduct our analyses\footnote{\url{https://github.com/CPJKU/musical_distribution_shift}}. 

The remainder of this article is structured as follows: Section \ref{sec:related} provides an overview of related works, including the AMT systems being evaluated; Section \ref{sec:method} introduces our new evaluation dataset and the applied metrics; Section \ref{sec:eval} presents the main findings, followed by an afterthought in Section \ref{sec:dds}. The article closes with takeaways and future directions in Section~\ref{sec:conclude}. An appendix \ref{sec:data} offers an additional statistical analysis of our new data from a musicological perspective.

\section{Related Work}\label{sec:related}

In recent years, advancements in Automatic Music Transcription (AMT) have been consistently driven by the application of Deep Neural Network (DNN) architectures, which would typically be trained to carry out the task in an end-to-end fashion \cite{bock2012polyphonic,sigtia2016endtoend,kelz2016potential,hawthorne2018onsets,kwon2020polyphonic,kong2021highresolution,hawthorne2021sequence,toyama2023automatic}.
Earlier models (similarly to preceding non-DNN-based AMT systems \cite{bay2009evaluation}) were limited to predicting note activity at the frame level.
A subsequent wave of improvements was brought about by a shift of focus towards explicit modelling and prediction of note onset and offset events \cite{hawthorne2018onsets,kwon2020polyphonic,kong2021highresolution,hawthorne2021sequence,toyama2023automatic}.

One common limitation of all these systems, however, is their dependence on a large, time-aligned, note-labelled training corpus of music, for learning to detect notes in polyphonic audio mixtures in a supervised fashion.
The datasets suitable for this purpose have been dominated by classical piano music: the most prominent examples are the SMD \cite{muller2011saarland}, MAPS \cite{emiya2010multipitch,emiya2010maps}, and MAESTRO \cite{hawthorne2019enabling} datasets.

Previous research has highlighted a tendency of these transcription systems to simply memorize frequently encountered note combinations, limiting their capacity to identify novel ones -- a phenomenon called the \textit{entanglement problem} in \cite{kelz2017experimental}.
More recent work has confirmed the persistence of this issue, now observed on a broader scale and described as \textit{corpus bias} in \cite{martak2022balancing}.
One possible way to view these results is as signs of sensitivity of these systems to \textit{musical distribution shift}.

Additionally, overfitting on sound-specific characteristics of the training data appears to reduce the reliability of these systems as well, specifically when applied to OOD recordings \cite{hawthorne2019enabling,edwards2024datadriven,hu2024musically}. While data augmentation techniques seem to alleviate this issue to certain extent, these gains seem to come at the expense of reduced ID performance \cite{hawthorne2019enabling,edwards2024datadriven, teney2023id}.
These results highlight the brittleness of current AMT systems in the presence of \textit{sound distribution shift}.

While datasets like SMD \cite{muller2011saarland}, SMD-Synth \cite{taenzer2021SMDsynth}, MAPS \cite{emiya2010maps}, or Studio-MAESTRO \cite{edwards2024datadriven} can all be used to measure the effects of particular shift in sound away from the MAESTRO data, the genre of their content is limited to Classical music. The IDMT-PIANO-MM dataset \cite{abesser2021benchmark}, while containing a small sample of Classical and Jazz pieces, is primarily designed to benchmark the effects of changes in sound, as it provides structured and well documented variations in piano instrument, room acoustics, and recording conditions.

\paragraph{Systems under Evaluation}\label{sec:related_systems}
In this work, we assess the major recent state-of-the-art DNN-based AMT systems, all of which were trained solely on classical piano music, primarily sourced from the MAESTRO dataset \cite{hawthorne2019enabling}.

\begin{itemize}
	
	\item \textbf{Onsets and Frames (OaF)} \cite{hawthorne2018onsets} -- a neural network architecture leveraging the concept of multi-task learning, using multiple heads trained with separate objectives to predict note onsets and frames (in subsequent updates also offsets and velocities) with specific information flows among the individual heads. Each head stacks CNN and RNN blocks.
	
	\item \textbf{High-Resolution Piano Transcription (Kong)} \cite{kong2021highresolution} -- a smart reformulation of the learning task through producing labels with higher temporal fidelity. The model learns to regress the continuous values of note onset and offset times, surpassing the limited temporal resolution of frame-wise input representation. This system additionally learns to model the activity of the piano sustain pedal.
	
	\item \textbf{Text-To-Text Transfer Transformer (T5)} \cite{hawthorne2021sequence} -- a generic sequence-to-sequence Transformer DNN architecture, trained on pairs of audio and MIDI for piano transcription in a supervised fashion, using the MAESTRO dataset. %
	
	\item \textbf{Hierarchical Frequency-Time Transformer (Toyama)} \cite{toyama2023automatic} -- a novel Transformer architecture highly tailored to work with spectrogram inputs.
	
	\item \textbf{Robustness-oriented re-training of the Kong model (Edwards)} \cite{edwards2024datadriven} -- the high-resolution architecture of the Kong model \cite{kong2021highresolution}, trained from scratch with various data augmentations, in an effort to increase robustness to sound distribution shift.
	
\end{itemize}

For brevity and convenience, we will refer to these systems as OaF \cite{hawthorne2018onsets}, Kong \cite{kong2021highresolution}, T5 \cite{hawthorne2021sequence}, Toyama \cite{toyama2023automatic}, and Edwards \cite{edwards2024datadriven} throughout this manuscript. %
We evaluate the later version of OaF, trained on the MAESTRO dataset, as documented in \cite{hawthorne2019enabling}.

\begin{figure}[t]
	\centering
	\includegraphics[width=\textwidth]{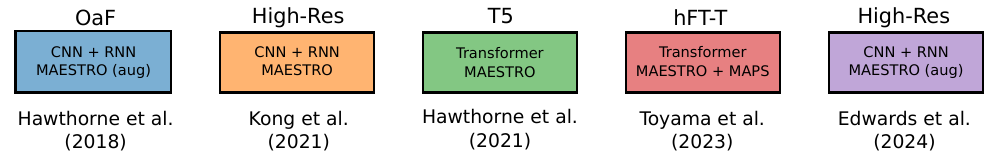}
	\caption{Systems under evaluation: abbreviated model architecture names (top); basic architectural types and training data sources (boxes); and the first publication authors (bottom). Note that both OaF and Edwards trained with sound-related data augmentations. The color scheme will be consistently followed later in various comparative result plots.}
	\label{fig:inkscape_models_legend_aug}
\end{figure}

\section{Methodology}
\label{sec:method}

To facilitate analysis of how AMT systems respond to different distribution shifts, we constructed a new dataset by first assembling MIDI files, and then producing audio recordings from them using a real piano, thereby generating both the audio data and the corresponding ground truth (note-level aligned labels).
The dataset comprises of three distinct collections of fine-aligned audio-MIDI pairs: (1) \textit{Genre}, (2) \textit{Random}, and (3) \textit{MAEtest} --- a $\sim$ 10\% subset of the MAESTRO test set.

The \textit{Genre} set collection was curated to represent 10 musical genres, including classical, hand-picked to maximize diversity. The curation process is described in more detail in Section \ref{sec:curating} below.

The \textit{Random} set was synthesized as entirely randomized note sequences, designed to emulate \emph{extreme} distribution shift, far exceeding what generally would be regarded as \emph{musical} by human standards, to examine the corpus bias under extreme conditions that deviate drastically from conventional tonal contexts.
This will enable probing the limits of existing AMT systems, effectively testing their robustness when confronted with highly unconventional inputs.
Additionally, by sampling the low-level musical parameters -- such as note pitches, onset times, durations, and velocities -- from distributions arbitrarily ``distant'' from musical ones (e.g., uniform or beta), we aim to provide an up-to-date assessment of the entanglement and corpus bias problems.
While this dataset is not adversarial in the traditional sense, its artificially challenging nature offers valuable insights into the performance of AMT systems under extreme and non-musical conditions.

Thirdly, we create the \textit{MAEtest} set, by rendering to audio 20 pieces from the MAESTRO test set, again using our Disklavier.
\footnote{Re-playing and recording the entire MAESTRO dataset on our mechanical piano was out of the question, for practical reasons.}
This will give us reference and calibration points regarding how our re-executions of the various systems compare to the respective results (on MAESTRO) that were reported in the original papers.
It will also -- in combination with the original MAESTRO recordings -- allow us to isolate and directly assess the impact of sound on the evaluated AMT systems.

\begin{figure}[t]
	\centering
	\includegraphics[width=\textwidth]{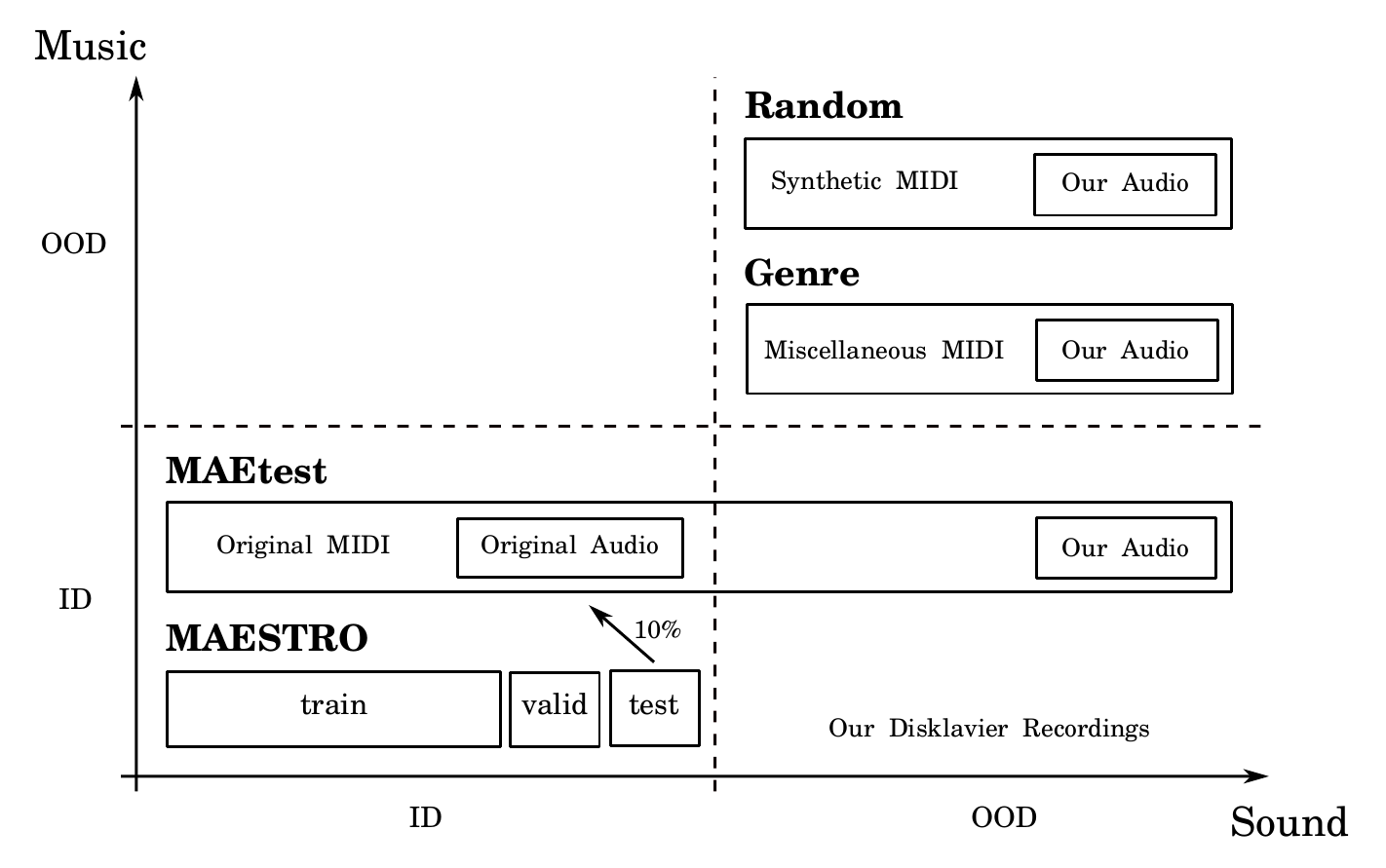}
	\caption{The data distribution situation in our used datasets, in terms of sound and music distribution shifts (ID / OOD).}
	\label{fig:inkscape_data_situ}
\end{figure}

As the experimental setup in terms of train/test ID/OOD data is somewhat complex, we provide a visual overview in Figure \ref{fig:inkscape_data_situ} and Table \ref{tab:data_situation}.
Overall, our strategy is as follows:
generally, we evaluate the five AMT models as they were trained by the respective authors, using both the inference source
code and the checkpoints with trained model parameters provided by them.
All five systems were trained on classical piano music, primarily sourced from MAESTRO (see also once again the concise summary in Fig.~\ref{fig:inkscape_models_legend_aug}).
The MAEtest set (which were not used for training)
consists of a set of ground truth MIDI files along with two sets of matching audio recordings -- the original audio provided by MAESTRO, and our own Disklavier recordings.
Since the musical material is classical pieces from MAESTRO, it follows the training distribution of these systems (i.e. \emph{music} is ID).
This data permits us to measure the particular impact of \emph{sound} distribution shift from MAESTRO (ID) to our Disklavier (OOD).

Once the impact of the fixed sound difference is accounted for, we can then measure the isolated effects of \emph{music} distribution shift from the classical music in MAEtest (ID) to music of various genres in our Genre set (OOD).
Our Random set will then permit us to go a step further, allowing to assess the impact of \textit{extreme} music distribution shift, all the way towards completely random note sequences (\textit{far} OOD).
Based on its intended utility, we refer to this new evaluation corpus -- MAEtest+Genre+Random -- as the \textit{Musical Distribution Shift (MDS)} Dataset.

\begin{table}[t]
	\caption{The distribution shifts of our data relative to the evaluated AMT systems}\label{tab:data_situation}
	\centering
	\begin{tabular}{@{}l|cc|cc@{}}
		\toprule
		\textbf{Dataset} & \multicolumn{2}{c|}{\textbf{Origins of the Data}} & \multicolumn{2}{c}{\textbf{Distribution Shift}} \\
		& MIDI & Audio  & Music & Sound \\ \midrule
		(1) Genre                    & Miscellaneous & Our Disklavier & \textbf{$\checkmark$} & \textbf{$\checkmark$} \\
		(2) Random                   & Synthetic & Our Disklavier & \textbf{$\checkmark$} & \textbf{$\checkmark$} \\ 
		\multirow{2}{*}{(3) MAEtest} & MAESTRO   & Our Disklavier        & \textbf{$\times$}     & \textbf{$\checkmark$}     \\
		& MAESTRO  & MAESTRO & \textbf{$\times$}     & \textbf{$\times$} \\
		\bottomrule
	\end{tabular}
\end{table}

\subsection{Curating the MDS Dataset}
\label{sec:curating}

\paragraph{Collecting the MIDI files for the Genre set}

The pieces for the Genre set are sourced from the ADL Piano MIDI dataset\footnote{\url{https://github.com/lucasnfe/adl-piano-midi/}} -- a collection of piano MIDI files created by combining existing MIDI collections and scraping the Internet, for the purposes of developing DNN-based emotion-conditioned music generation systems \cite{ferreira2020computer}.
It is our resource of choice due to being a piano-specific MIDI collection with the largest genre diversity from among what is to be found freely available on the Internet.

We hand-picked the following 10 genres to maximize diversity, coverage of the space, sensibility of genre for solo piano performance, and favoring genres represented by larger numbers of MIDI files: Classical, Blues, Rock, Soul, Jazz, Latin, Pop, Country, Folk, and World.
To reduce the chance of having low quality or unrepresentative pieces selected, we apply filtering based on the following markers: \\

\noindent
- exclude pieces lasting outside the time interval of 2 to 4 minutes (120 - 240 seconds) \\
- exclude pieces totalling over five seconds of gaps (absence of note activity) \\
- exclude pieces with average polyphony level below 3 \\
- exclude pieces using notes with fewer than 20 unique velocity values \\
- exclude pieces using notes with fewer than 15 unique pitch values \\
- exclude pieces containing any non-piano instruments \\
- exclude pieces containing pitch bend commands \\

\begin{figure}[t]
	\centering
	\includegraphics[width=\textwidth]{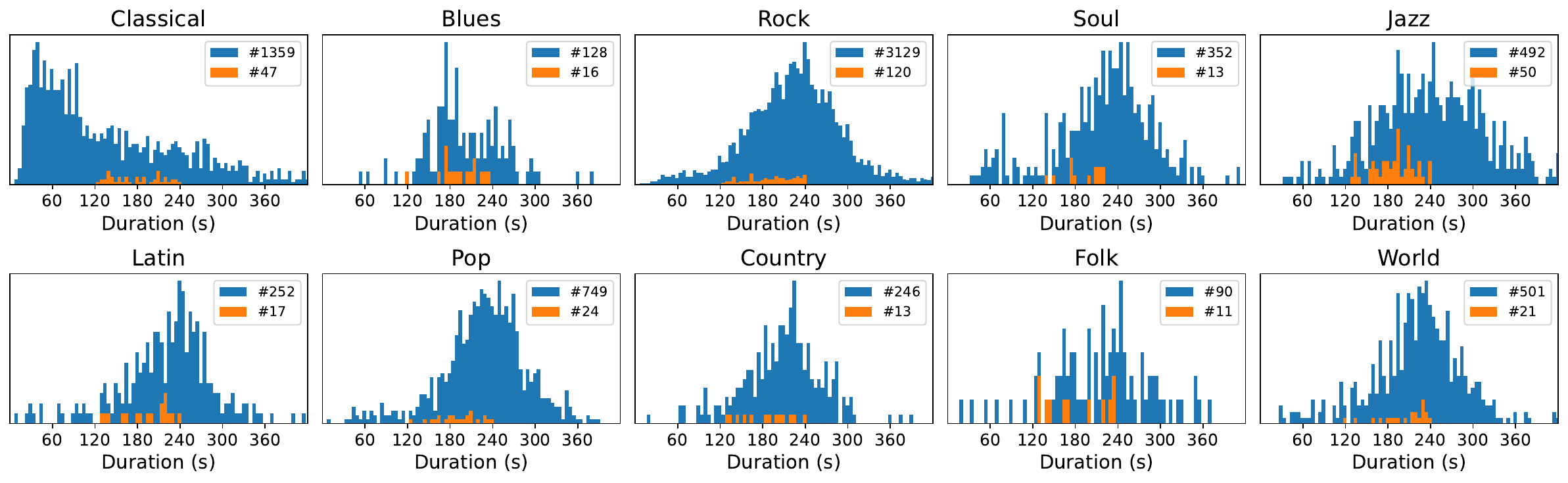}
	\caption{Counts of MIDI files eligible for selection: before (blue) and after (orange) our filtering.}
	\label{fig:genre_counts_filter_impact}
\end{figure}

We found and adjusted these thresholds empirically, while trying to eliminate repeatedly occurring sample pieces of questionable quality, completeness, or genre representativeness.\footnote{We often ran into pieces that seemed to contain only a single hand performance, or a single part of an ensemble composition, and the most important filter to get rid of those ``incomplete'' performances seemed to be the average polyphony level filter.}
After constraining the piece candidates using these exclusion criteria, we pseudo-randomly sampled 10 unique pieces (MIDI performances) for each of the 10 genres.\footnote{In the selected MIDI tracks that contained more than 1 instrument, we merged the instruments and handled any note collisions by adjusting the onset and offset times of the collided notes, to eliminate their overlap while preserving the note's total active time -- treating those as note re-onsets.}
Figure \ref{fig:genre_counts_filter_impact} illustrates the strong impact of our candidate filtering, highlighting the scarcity of quality piano MIDI data with genre annotations.

\paragraph{Synthesis of MIDI files for the Random set}

The Random set is composed of 72 ($24 \times 3$) sequences, each lasting 2 minutes in length, systematically varying in degree of polyphony and dynamics.
The polyphony ($p$) ranges from $1$ to $24$ simultaneous voices, while the dynamics range ($d$) is categorized into three levels: narrow (0: $60-68$ [vel]), medium (1: $32-96$ [vel]), and wide (2: $1-127$ [vel]); each prescribing a range of allowed MIDI velocity values.

Each sequence is populated with $p$ parallel streams of notes, featuring MIDI pitch values randomly sampled from a uniform distribution $\sim U(21, 108)$, covering the entire range of piano claviature (88 keys).
The velocities are uniformly distributed within the range specified by the dynamics level $\sim U(d_{\min},d_{\max})$.
The note durations are drawn from a Beta distribution $Beta(\alpha=2,\beta=5)$ spanning $0.01$ to $5.00$ seconds.
We prefer the Beta distribution, because it can resemble the note duration statistics of real music -- see the histogram of note durations in Figure \ref{fig:datastats_midi_all_juxtaposed} for a visual comparison -- without trying too hard to match its over-representation of short (staccato) notes.

Furthermore, each note additionally undergoes slight trimming at both ends, randomly having its duration reduced by up to 1\% on each end.
This introduces slight discontinuities in the voices without drastically compromising the maintained polyphony.
Our aim with this was to at least partially eliminate the otherwise perfect temporal correlations of offsets and onsets of the voice-adjacent notes, introducing further randomization while reducing the unrealistic uniformity of the voices.
As a side effect, these tiny gaps in the parallel voices introduce momentary drops in the maintained polyphony level. Co-occurrence of these gaps compounds with growing prescribed polyphony level $p$.

While the MAPS dataset contains a RAND subset comprising of random chords \cite{emiya2010maps}, the notes therein are fully synchronous in time, and their loudness only has 2 categories: homogeneous ($[60-68]$) and heterogeneous ($[32-96]$).
Our Random set additionally randomizes and de-correlates onsets and durations of notes, reaching further away from the statistical structures found in human-composed music.
Our additional inclusion of higher-than-practical polyphony degrees and an expanded range of dynamics levels results in a more exhaustive and varied experimental benchmark for probing the limits of future AMT systems.
For qualitative illustration, Figure \ref{fig:random_dynamics_showcase} (Appendix \ref{sec:data}) shows samples produced by our procedure.

\paragraph{Rendering the pieces and recording audio files}

To reduce any possible influence of variations in sound and isolate the effects of musical distribution shift, we build our corpus by recording automated performances of the MIDI targets using a Yamaha Disklavier\footnote{\url{https://usa.yamaha.com/products/musical_instruments/pianos/disklavier/index.html}} grand piano.

\begin{figure}[t]
	\centering
	\includegraphics[width=0.66\textwidth]{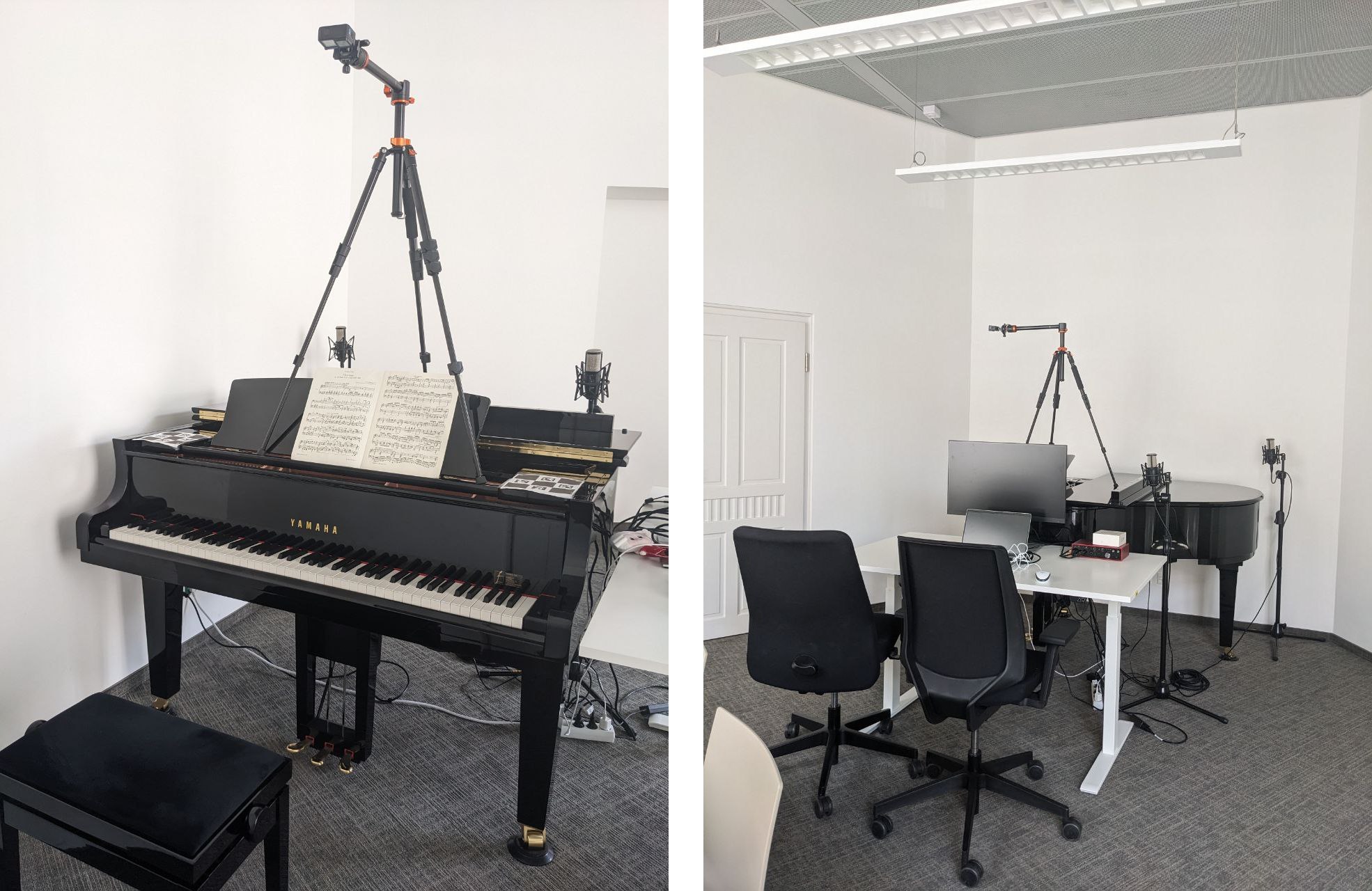}
	\caption{Our Disklavier piano recording setup.}
	\label{fig:psk_disklavier_setup}
\end{figure}

We produce our MIDI-synchronized audio recordings using the Python package \texttt{piano-capture}\footnote{\url{https://github.com/almostimplemented/piano-capture}} to play the MIDI files on a Yamaha Disklavier Enspire ST C1X.
The recordings are captured with a Focusrite Scarlett 18i8 audio interface and a pair of AKG P420 microphones in a moderately bright, fully carpeted room with asymmetric geometry and minimal background noise, simulating studio-like conditions (see Figure \ref{fig:psk_disklavier_setup}), same as in \cite{hu2024musically}.
This ensures the sound-related aspects of our data generating distribution to be the same for all of the recorded audio.

\subsection{Assessing the MDS Dataset}

\begin{figure}[t]
	\centering
	\includegraphics[width=\textwidth]{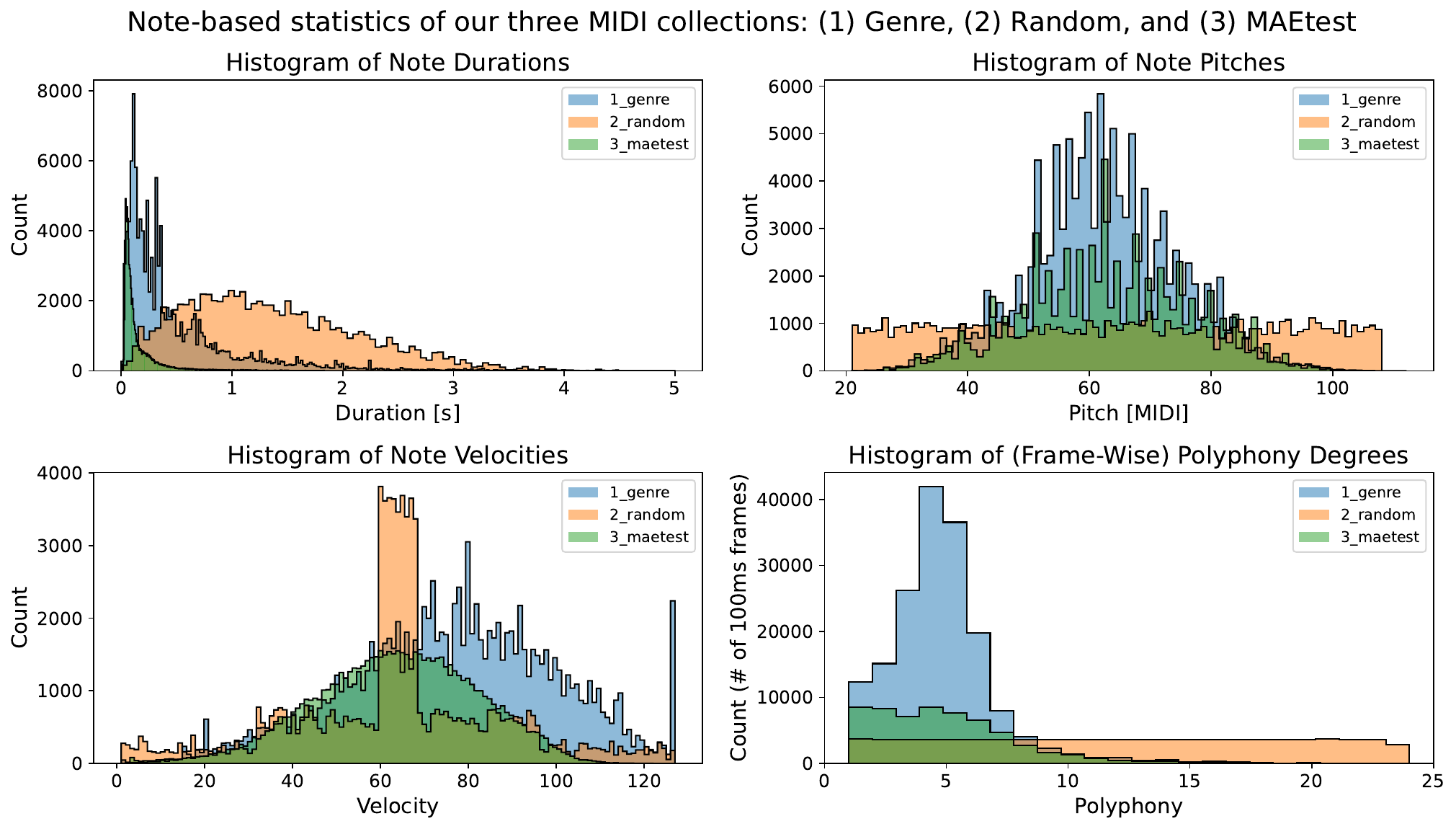}
	\caption{Comparison of our 3 collections in terms of basic note-level statistics: note durations (top left); pitch (top right); MIDI velocity (bottom left); polyphony level (frame by frame; bottom right).}
	\label{fig:datastats_midi_all_juxtaposed}
\end{figure}

Ultimately, these three collections of pieces will provide the basis for our analysis of SotA AMT systems. A comparison of their low level MIDI note-based statistics is shown in Figure \ref{fig:datastats_midi_all_juxtaposed}.
Following our design choices, the Random set unsurprisingly exhibits strong deviations from the other two collections in all statistics, materializing our intended extremities in music distribution shift.

We surmise that the observable differences between our Genre set and the MAEtest collection -- such as smoothness of the velocity statistics -- might be related to the following two factors:
\begin{enumerate}
	\item First, the MIDI capture fidelity of a piano performance (along with the performers' skills) may wildly vary among the Genre pieces found on the Internet, compared to the high capture fidelity (and performers' skill) that are de facto guaranteed for all MAESTRO performances.
	\item Moreover, the often superior complexity of classical music compared to most genres might lead to more evenly distributed note statistics, especially for their dynamics and polyphony.
\end{enumerate}

\begin{table}[t]
	\centering
	\caption{Size comparison of our evaluation subsets (MIDI).}\label{tab:midiset_sizes}
	\begin{tabular}{@{}lrrr@{}}
		\toprule
		\textbf{MIDI Set} & \textbf{Hours} & \textbf{Notes} & \textbf{Pieces} \\
		\midrule
		(1) Genre          & 5.25           & 118,388         & 100           \\
		(2) Random         & 2.40           & 79,054          & 72            \\
		(3) MAEtest        & 1.84           & 73,060          & 20            \\
		\bottomrule
	\end{tabular}
\end{table}

Accompanying the statistics in Figure \ref{fig:datastats_midi_all_juxtaposed}, Table \ref{tab:midiset_sizes} documents the sizes of our three subsets.
In the light of today's large data regimes -- such as when training foundation models \cite{gardner2023llark,ma2024foundation} -- these might appear as ``small''. 
Since our data is not intended for training, but rather evaluation, the sample sizes should be sufficient in contributing new ``signal'' towards benchmarking OOD performance of the evaluated AMT systems.
We kept the data size moderate due to both the scarcity of readily-available high quality performances with balanced coverage of genres (Figure \ref{fig:genre_counts_filter_impact}), and the time complexity of data acquisition -- setting up and recording pieces on a Disklavier is a much more tedious process than rendering audio via piano samplers, and is bounded by a speed of $1 \times$ real-time.

Further statistical analysis of our new collection -- focused mainly on the Genre set -- can be found in the dataset report, attached as appendix \ref{sec:data}.
It offers a detailed look into the statistical and musical properties of the Genre and Random subsets. For the Genre subset, we find that most genres exhibit similar distributions in pitch, pitch class, and velocity, but show greater variability (both within- and between-group) in note density and polyphony levels. Regarding harmony and tonality, with the exception of Jazz and Blues, most genres display a higher frequency of consonant intervals -- particularly thirds -- indicating a predominantly tonal musical structure across the dataset.

\subsection{Evaluation Metrics}
\label{sec:metrics}
The Music Information Retrieval (MIR) community has a history of evaluating AMT systems using various quantitative metrics \cite{bay2009evaluation}, mostly evolved around the distinction between two dominant output representations of polyphonic music transcription systems:
(a) \textit{frame-level} -- where the activity of notes is predicted independently for each individual time frame in a grid given by a regularly spaced temporal quantization of the signal (such as, e.g., resulting from STFT analysis); %
(b) \textit{note-level} -- where activity of notes is predicted at the level of note events; this is generally called the Note Tracking task in the literature.
The most common AMT evaluation metrics in MIR, both at the frame and note levels, are the basic Information Retrieval (IR) measures \textit{recall}, \textit{precision}, and \textit{F1-score}.
We naturally start our analysis with these standard IR metrics, to make our results directly comparable to others in the literature. %
Beyond this, however, we will additionally apply a set of recently proposed, \textit{musically informed} metrics, hoping to draw additional (and more musically relevant) insights from these.

\subsubsection{Information Retrieval Metrics}\label{sec:metrics_ir}

The standard IR-based evaluation treats the frame- and note-level predictions the same way as documents retrieved by IR systems are treated in the IR literature.

\paragraph{Frame-level}
At the level of frame-wise predictions of note activity, two piano-roll-style binary matrices of equal shape (the predicted one, and the ground truth) are compared, where each $[\text{pitch} \times \text{time}]$ cell is considered a `retrieval event', yielding one of the matched (true positive/negative) or mismatched (false positive/negative) options.

\paragraph{Note-level}
At the level of note-wise predictions, two (not necessarily equally large) collections of notes are compared and the IR metrics are computed with notes playing the role of documents in the traditional IR setting.
As musical notes are nuanced objects defined by their specific properties -- [pitch, onset, offset, (optional) MIDI velocity (related to loudness)] -- there are different ways of comparing them to each other for the purpose of determining what constitutes a match (true positive).
This fact has given rise to several variations of the note-level IR metrics.
The three most popular ones that became \textit{de facto} standard for modern AMT system benchmarking are progressively composed based on the following match conditions:

\begin{itemize}
	\item [] \textbf{Note with Onset:}
	predicted pitch lies within +/- 50 cents of the reference note's pitch; predicted onset time lies within +/- 50 ms of the reference note's onset.
	\item [] \textbf{Note with Onset \& Offset:}
	same conditions as Note with Onset, plus:
	predicted offset time lies within +/- max(50 ms, 20\% of duration) of the reference note's offset.
	\item [] \textbf{Note with Onset \& Offset \& Velocity:}
	same conditions as Note with Onset \& Offset, plus:
	predicted velocity lies within 10\% distance\footnote{The threshold of \texttt{0.1} is used after global re-scaling of the estimated velocities, see \cite{hawthorne2018onsets} for the details.} of the reference note's velocity.
\end{itemize}
To decouple velocity estimation from offset detection, we additionally quantify the following variant:
\begin{itemize}
	\item [] \textbf{Note with Onset \& Velocity:}
	same conditions as Note with Onset, plus:
	predicted velocity lies within 10\% distance of the reference note's velocity.
\end{itemize}

\subsubsection{Musically Informed Metrics}\label{sec:metrics_mi}

The standard IR metrics have driven progress in AMT by quantifying low-level performance, but they are generic and blind to music-specific effects of transcription imperfections. 
We thus complement them with a set of `musically informed' piano transcription metrics that have recently been proposed \cite{hu2024musically}, to provide a more nuanced assessment. 
While measures like precision, recall, and F1 score offer valuable insights into note-level accuracy, these musically informed metrics delve into aspects related to music performance and our perception of it, such as dynamics, articulation, rhythmic precision, and harmony-related aspects, all of which are crucial for assessing the musical fidelity of transcriptions.
In this way, we aim to evaluate how well a transcription system captures expressive qualities of piano performances, including variations in loudness, note duration, and temporal alignment. 
The idea is to ensure that the transcription results align more closely with human musical perception, offering a more comprehensive evaluation framework for piano transcription systems.
The metrics proposed in \cite{hu2024musically} are the following:

\paragraph{Expressive Timing.} Expressive timing in solo piano performance denotes micro-timing deviations in note onsets relative to the metrical grid of the score. It is commonly measured using the \textit{Inter-Onset-Interval (IOI)}, that is, the amount of time passed between two consecutive notes belonging to the same stream \cite{repp1994relational, desain1994does} \footnote{We use the term \textit{stream} as a generalization of the concept of a voice in polyphonic music~\cite{temperley:2009jnmr}.} To evaluate whether a transcription accurately preserves micro-timing deviations, both the transcription and the ground truth are separated into streams, and the correlation between their respective streamwise IOI sequences is measured.

\paragraph{Articulation.} Articulation in piano performance refers to how notes are connected or separated, resulting in expressive strategies such as legato, staccato, or marcato. Computationally, articulation is measured using the \textit{Key-Onset-Ratio (KOR)}, which is the ratio between the offset-to-onset interval and the onset-to-onset interval of consecutive notes \cite{Bernays2014investigating}. To evaluate how accurately a transcription captures expressive articulation, the streamwise KOR sequences of the ground truth and transcribed performance are compared using Pearson correlation.

\paragraph{Harmony.} Harmonic tension is a critical factor in expressive piano performance, influencing decisions related to tempo and dynamics \cite{cancino2018computational, herremans2019towards}. To evaluate the preservation of harmonic tension in a transcription, we use three harmonic features derived from Chew’s spiral array model \cite{chew2016playing}. This model represents pitch classes, chords, and keys in a three-dimensional space, where spatial proximity reflects tonal relationship.

The first metric, \textit{Cloud Diameter}, measures the maximal tonal spread among notes in a musical segment, while the second, \textit{Cloud Momentum}, captures the harmonic movement by measuring tonal distance between consecutive sections. The third metric, \textit{Tensile strain}, quantifies the relative tonal distance between the current segment and the overall harmonic center across the piece. We compute these metrics on overlapping windows for both the original and transcribed MIDI, then analyze their correlations to evaluate how well harmonic tension is maintained. Note that these harmony metrics primarily reflect the presence or absence of pitches rather than timing accuracy of note onsets or offsets.

\paragraph{Dynamics.} In addition to temporal and pitch/harmony aspects, expressive dynamics are a key feature of expressive performance. To evaluate how well a transcription captures dynamics, the loudness ratio between the melody and bass lines is used as a proxy, reflecting the transcription's preservation of the performance's dynamic range. Loudness is estimated as the `energy' of each stream (melody or bass) based on MIDI velocity, following Dannenberg’s model \cite{dannenberg2006interpretation}. 
Unlike conventional IR metrics, this measure focuses exclusively on the dynamics and energy captured in the transcription, by comparing the loudness ratios of the ground truth and transcribed performance MIDI and computing the correlation between these.

\section{Experiments, Evaluation, and Analysis}\label{sec:eval}

To assess our selection of AMT systems (see Section \ref{sec:related}), we use both the inference source code and the checkpoints with trained model parameters as provided by the respective authors.
We run these systems to transcribe all the relevant audio files, and keep the piece-wise transcriptions as MIDI files.
The individual transcription files are then paired with corresponding ground truth MIDI files for computation of various performance metrics.
This leads to an evaluation at the piece level, yielding a large grid of numerical results with a single value for each unique combination of \texttt{[piece, system, metric]}.
In the following, we report our results not at the level of individual pieces, but rather as various aggregations along different dimensions of our data, and discuss the most relevant observations.

\subsection{A First High-level View}
\label{sec:high_level}

\begin{figure}[bp]
	\centering
	\includegraphics[width=\textwidth]{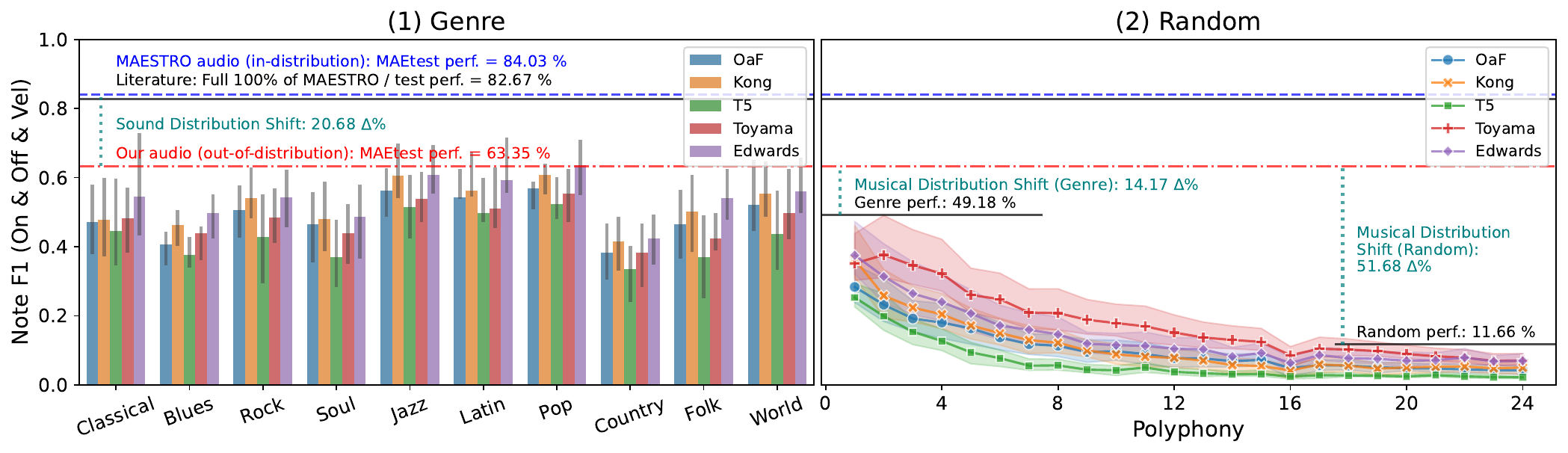}
	\caption{Performance of the evaluated AMT systems on our MDS Dataset in terms of the most nuanced IR metric: Note-level F1-score with Onset, Offset and Velocity.}
	\label{fig:eval_all_note_on_off_vel}
\end{figure}

Let us start the analysis with an attempt at quantifying the overall effects of sound and music distribution shifts, and the effect of using our smaller MAEtest set as a representative subsample of the full MAESTRO test set.
Figure \ref{fig:eval_all_note_on_off_vel} displays the aggregate inference performance, as measured by \textit{Note-level F1-score with Onset, Offset, and Velocity}, of the five AMT systems on all four evaluation subsets -- (1) Genre, (2) Random, and the two audio versions (Disklavier and MAESTRO original) of (3) MAEtest.
The data bars on the left show model-genre performance, averaged over the pieces in each genre in our (1) Genre set.
The line plots on the right side display model performance as a function of polyphony degree, averaged across the three levels of dynamics in our (2) Random set.
The error bars on the left as well as the data bands on the right represent inter-quartile ranges (50\% percentile intervals).

The black solid horizontal line through both subplots serves as an approximate\footnote{Edwards et al. \cite{edwards2024datadriven} do not report their model's performance in this metric.} reference for the average performance of the five AMT systems on the full test set of MAESTRO \cite{hawthorne2019enabling}, based on the numbers reported in the respective original publications.
The blue dashed line reports the average model performance as measured by transcriptions of our (3) MAEtest subset ($\sim$10\% of MAESTRO original audio).
The red dash-dotted line further quantifies the model's average performance on our \textit{Disklavier recordings} of the MAEtest subset.
The teal dotted vertical lines, finally, depict the differences in average model performance attributable to the different respective kinds of data distribution shift between individual subsets of our data. 
Generally it can be observed that AMT models do not perform consistently well under varying musical and acoustic conditions. Our four evaluation subsets can help in disentangling and quantifying the effect of each respective distribution shift.

We start by observing the close proximity of the blue dashed line to the black line. We take this as reassuring evidence that (a) validates the correctness of our experimental reproduction of the AMT systems and (b) confirms that MAEtest is sufficiently representative of the full MAESTRO test set.
The gap between the red and blue lines depicts the difference in performance that is attributable to the specific \textit{sound distribution shift} between the MAESTRO and Disklavier recordings.

Next, even after accounting for the performance loss due to changing sound conditions, notable performance discrepancies remain, varying in magnitude depending on the musical content. These gaps, visible as gaps between the red dash-dotted line and the average model performance per genre (on the left) and per polyphony level (on the right side) highlight the role of the underlying \textit{musical material}.
On the right half of Figure \ref{fig:eval_all_note_on_off_vel} we additionally overlay the average performance across all models and data points -- from the (1) Genre set on the left, and (2) Random set on the right -- as short black lines with labels.
The shift within the musical domain (from MAEtest to Genre pieces), amounts to a transcription performance degradation of $\sim$ 14\%, while the shift towards the non-musical domain (from MAEtest to Random pieces) is noticeably larger, at $\sim$ 52\%.
In the following sections, we attempt to dissect the factors contributing to these types of distribution shift effects.

\begin{table}[t]
	\caption{Impacts of the three different distribution shifts, measured with our structured MDS Dataset as average model performance on the subsets in terms of all the considered Metrics. The reported values are differences ($\Delta$) between performance before and performance after distribution shift. Thus, as in Figure \ref{fig:eval_all_note_on_off_vel}, positive values reflect a \textit{decrease} in performance caused by the particular distribution shift, while negative values reflect the cases where a particular distribution shift even improved a metric. \textbf{Bold} highlights the largest impact within each metric sub-group.}
	\label{tab:all_dist_shifts}
	\centering
	\begin{tabular}{lrrr}
		\toprule
		\textbf{Metric} & \textbf{Sound [$\Delta$]} & \textbf{Music (Genre)[$\Delta$]} & \textbf{Music (Random)[$\Delta$]} \\
		\midrule
		Frame F1 & 8.33 & -6.57 & 26.99 \\
		\midrule
		Note F1 (On) & 2.43 & 1.62 & 20.46 \\
		Note F1 (On \& Off) & 16.02 & -1.99 & 47.79 \\
		Note F1 (On \& Vel) & 9.58 & \textbf{22.05} & 49.55 \\
		Note F1 (On \& Off \& Vel) & \textbf{20.68} & 14.17 & \textbf{51.68} \\
		\midrule
		Articulation (Melody KOR) & \textbf{11.33} & \textbf{4.05} & \textbf{23.17} \\
		Articulation (Bass KOR) & 10.83 & -0.98 & 14.02 \\
		\midrule
		Timing (Melody IOI) & 0.49 & \textbf{13.96} & \textbf{37.77} \\
		Timing (Accompaniment IOI) & \textbf{5.40} & 5.44 & 14.46 \\
		\midrule
		Harmony (Cloud Diameter) & -4.24 & 12.66 & \textbf{69.83} \\
		Harmony (Cloud Momentum) & \textbf{-9.65} & 15.88 & 22.43 \\
		Harmony (Tensile Strain) & -3.23 & \textbf{25.18} & 61.24 \\
		\midrule
		Dynamics (Loudness Ratio) & 9.87 & 11.51 & 20.00 \\
		\bottomrule
	\end{tabular}
\end{table}

Figure \ref{fig:eval_all_note_on_off_vel} is based on one particular evaluation metric (\textit{Note-level F1-score with Onset, Offset, and Velocity}). To give a complete picture, Table \ref{tab:all_dist_shifts} summarizes the effect sizes of the global distribution shifts in terms of all metrics in our suite, both IR (F1) and musically informed. The reported values represent the difference in performance (over all models) between (a) MAEtest on the MAESTRO and our Disklavier audio environment (sound shift); (b) Classical (MAEtest Disklavier) and average of all genres (Genre dataset Disklavier; genre-induced music distribution shift); and (c) Classical (MAEtest Disklavier) to Random Disklavier (extreme music distribution shift).
The values in the 5th row of the table -- Note F1 (On \& Off \& Vel) -- correspond to the teal labeled gaps in performance between subsets of the MDS Dataset in Figure \ref{fig:eval_all_note_on_off_vel}.
The other rows of this table complement this calculation with the use of the remaining metrics.

\subsection{Beyond Classical Music: The Genre Set}
In the following, we 
analyze the performance of the selected AMT models as measured on our Genre set. To better understand the impact of genre on model performance, we complement the note-level metrics with the musically informed ones. Then, we provide a comparative view on the effects of sound distribution introduced by a change in the recording environment, and musical distribution shift arising from the combined genre variations.

Genre labels in this work are applied heuristically (based on metadata from the source dataset \cite{ferreira2020computer}), acknowledging their fluid nature and the simplification inherent in piano renditions of originally genre-specific material \cite{fabbri1982theory}.
Despite the limitations of genre taxonomy, we observe measurable differences across genres, both in AMT performance and in their underlying musical characteristics. We provide a statistical analysis of Genre set in Appendix \ref{sec:data} as a supplementary step to contextualize these results.

\subsubsection{Performance on the Genre Set}
On the left side of Figure \ref{fig:eval_all_note_on_off_vel}, different-colored bars show the performance of our suite of AMT systems on genre-specific slices of the Genre set data.
All AMT models exhibit sensitivity to genre changes. While their absolute performance levels may differ, all systems exhibit similar relative patterns across genres, suggesting that genre-specific musical features consistently challenge each model. In other words, genre-induced shifts affect all systems in broadly comparable ways, indicating that these musical variations have a systematic influence on transcription performance in these particular AMT systems.

Perhaps counter-intuitively, the performance on the Classical genre is not closest to the MAEtest reference, as Jazz, Latin and Pop seem to show the smallest performance drop, while Country, Blues, Soul and Folk show the largest one.
The Classical genre seems to differ from the rest by how starkly the Edwards system outperforms the Kong one, but also by how closely matched the performance of the remaining four systems is within this genre, compared to others.

\paragraph{Note-level Breakdown}

\begin{figure}[t]
	\centering
	\includegraphics[width=\textwidth]{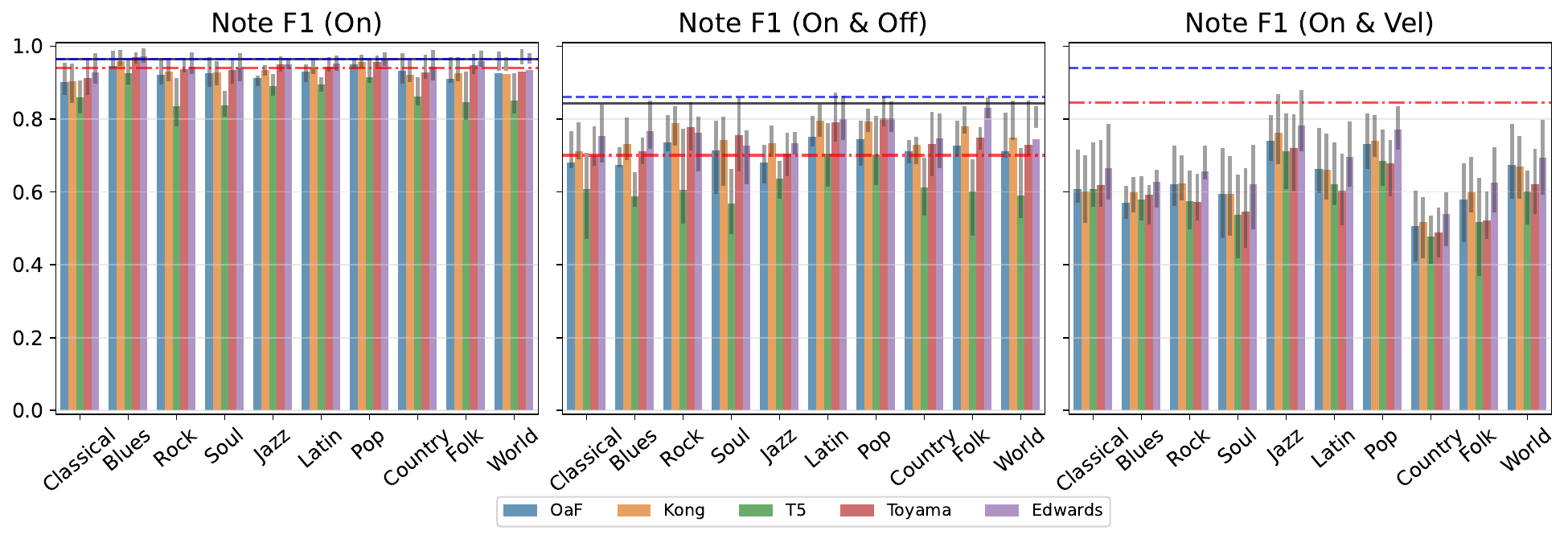}
	\caption{Performance evaluation on the Genre set in note-level metrics complementing the full Note-level variant reported in Figure \ref{fig:eval_all_note_on_off_vel}.}
	\label{fig:eval_genre_note_triplet_breakdown}
\end{figure}

Figure \ref{fig:eval_genre_note_triplet_breakdown} shows the Genre set performance evaluation, analogous to the left half of Figure \ref{fig:eval_all_note_on_off_vel}, using the other three Note-level F1-score metric variants described in Section \ref{sec:metrics_ir}.
The horizontal lines depicting performance levels on MAESTRO data follow the legend and semantics of Figure \ref{fig:eval_all_note_on_off_vel}.

The most simple metric variant -- Note with Onset only -- shows very small variation in performance both within and across different genres.
Adding the Offset detection condition (On \& Off) seems to increase variance in performance among models within individual genres, while cross-genre differences are only slightly more pronounced.
The global drop in performance (w.r.t. Onset only) might be largely due to sound, based on the corresponding MAEtest measurements.

On the other hand, the Velocity detection condition (On \& Vel) seems to introduce more pronounced differences between genres, larger performance drop compared to Onset \& Offset, and seems way less attributable to sound.
The sizes of error bars (showing inter-quartile ranges) indicate broad increase in performance variation, which suggests that on the piece level, the ``difficulty'' of velocity estimation is quite varied.

\paragraph{Musically Informed Analysis}

\begin{figure}[t]
	\centering
	\includegraphics[width=\textwidth]{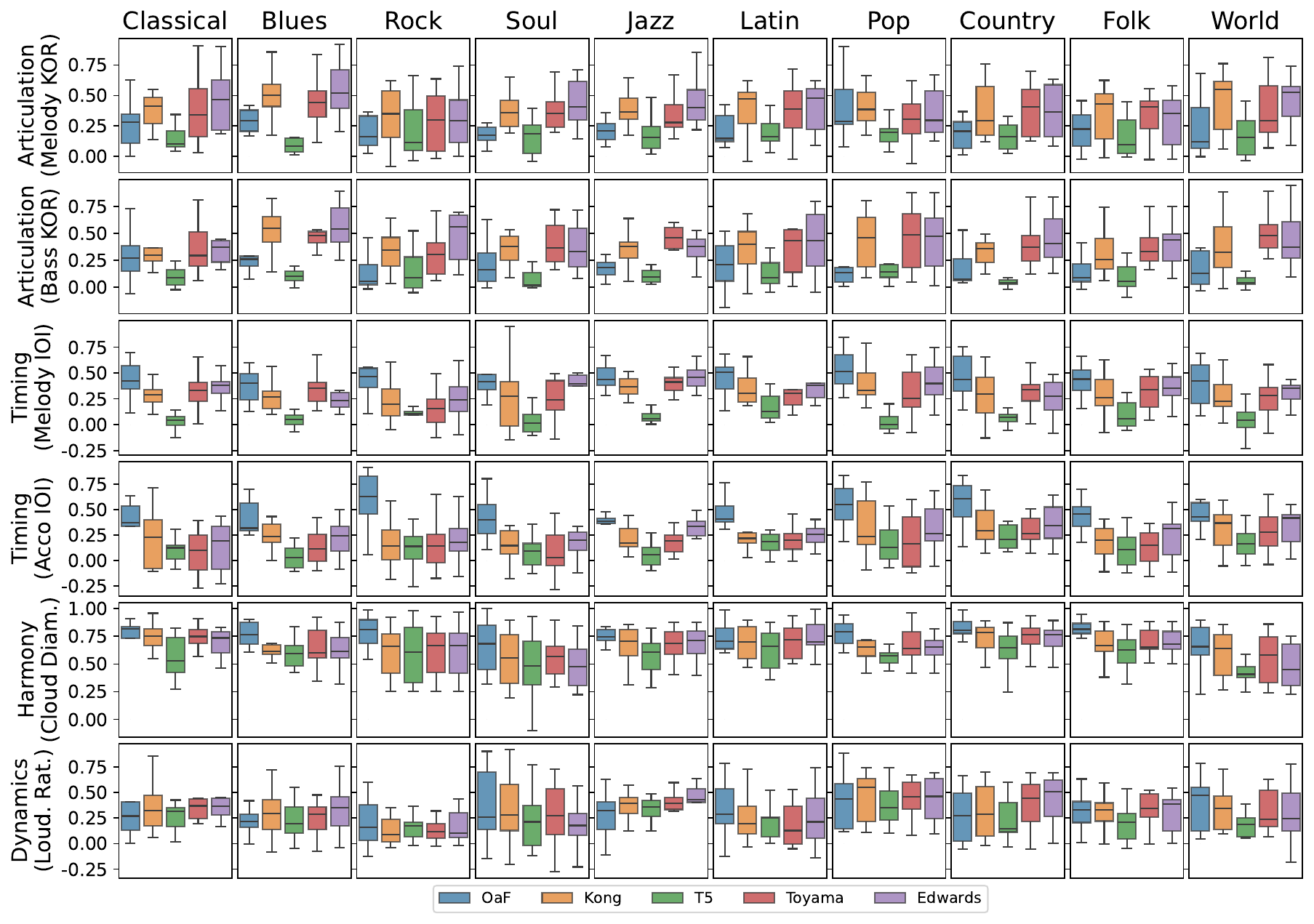}
	\caption{
		Performance on the Genre set in selected Musically Informed Metrics.
		Rows: metrics; columns: genres; color: models.
		The value axes are synchronized within rows for direct comparability.
	}
	\label{fig:eval_genre_MIMs}
\end{figure}

Figure~\ref{fig:eval_genre_MIMs} presents the performance per model and genre as evaluated on a selection of the \textit{Musically Informed Metrics}.
Looking at the performance as evaluated on the articulation and timing metrics, there is no common trend with respect to a particular genre being more difficult to transcribe for all models, as all models perform variably across genres in both the melody and bass streams. With respect to articulation, it seems that all models apart from T5 perform relatively well on the Classical genre, whereas the performance in other genres such as pop, Rock and World, exhibits more spread, as seen in Kong, Toyama and Edwards. It is interesting to compare Kong and Edwards on the bass stream articulation performance, as here it can be seen that a model generalized on the sound distribution (Edwards) exhibits, across some genres (i.e. Latin, Soul, Country) more variability in performance than without generalization (Kong).

With respect to the harmony-related metrics, we can see that the Cloud Diameter metric (comparing the tonal distance of notes measured in the reference and in the transcription) is rather low across all models in the Soul genre. Indeed, the musicological analysis of the pieces in the genre (see Appendix \ref{sec:data}, Figure~\ref{fig:genre_seventh}) reveals the highest proportion of the Minor Seventh chord in this genre, where it occurs more than double as often as, e.g., in the Classical genre. Most models exhibit considerable variability in the Rock genre, which features the highest proportion of the Dominant Seventh Chord, again occurring at roughly double the rate observed in Classical Music. 

Finally, evaluation of the dynamics metric suggests, that in some genres -- such as Classical, Jazz, and Folk -- it appears to be more consistently preserved, and in some cases (Jazz), also simpler than others (such as Country, Soul, and Pop). This observation is supported by Figure~\ref{fig:genre_vel}  in the Appendix, which shows that the latter three genres display a higher-centered velocity distribution compared to Classical (generally also visible in Figure~\ref{fig:datastats_midi_all_juxtaposed} as a difference between MAEtest and Genre sets).

\subsubsection{Genre-induced Distribution Shifts}
\label{sec:eval_genre_ds}

\begin{figure}[t]
	\centering
	\includegraphics[width=\textwidth]{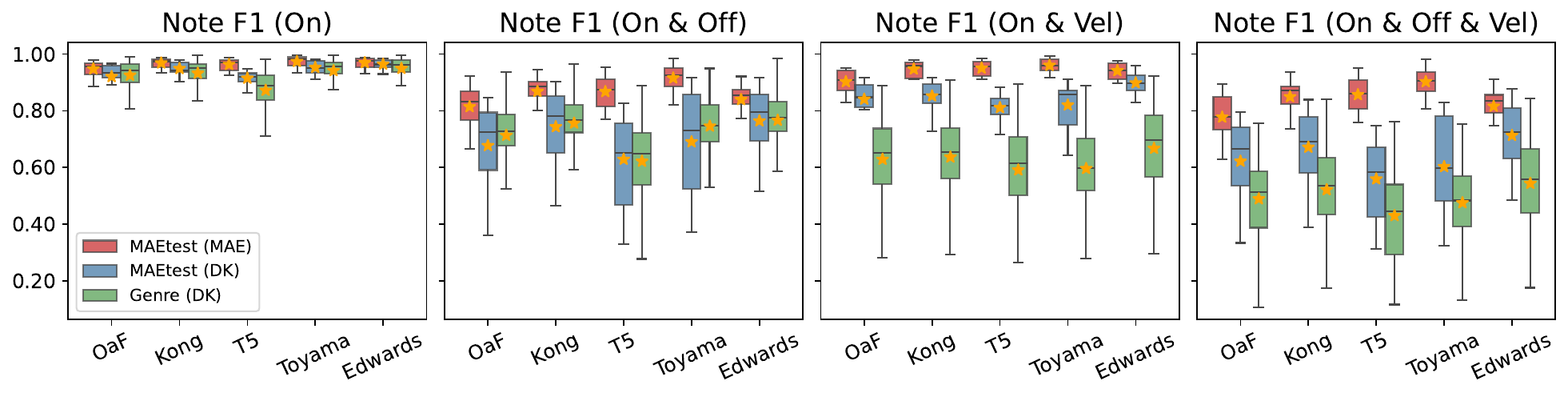}
	\caption{Note-level F1-scores for the Sound- and Genre-induced Distribution Shifts.}
	\label{fig:eval_sds_mds_combo_IRMs}
\end{figure}

\begin{figure}[t]
	\centering
	\includegraphics[width=\textwidth]{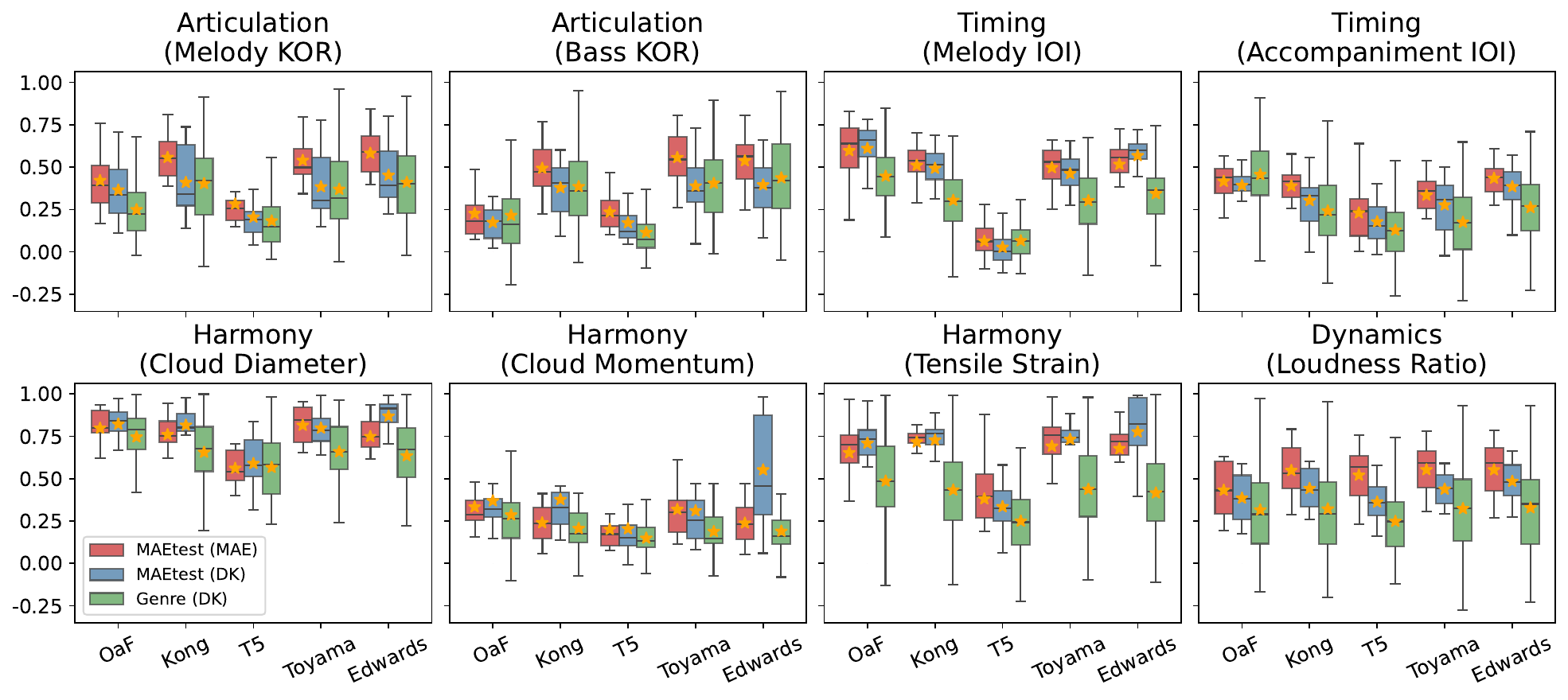}
	\caption{Musically Informed Metrics evaluation for the Sound- and Genre-induced Distribution Shifts.}
	\label{fig:eval_sds_mds_combo_MIMs}
\end{figure}

The first two columns of Table \ref{tab:all_dist_shifts} -- the model-averaged Disklavier-induced sound distribution shift and Genre-induced musical distribution shift -- can be further interpreted with the aid of Figures \ref{fig:eval_sds_mds_combo_IRMs} and \ref{fig:eval_sds_mds_combo_MIMs}: displaying triplets of performance statistics on the three relevant MDS subsets separately for each of the five AMT systems.
The differences between the first pair of candles (red vs. blue) reveal the impact of sound shift from MAESTRO audio to our Disklavier audio on the MAEtest pieces $\rightarrow$ 1st column in Table \ref{tab:all_dist_shifts}.
The differences between the second pair of candles (blue vs. green) show the additional musical shift from MAEtest pieces to the Genre pieces (both Disklavier) $\rightarrow$ 2nd column in Table \ref{tab:all_dist_shifts}.
Orange stars mark the means.%

The 2nd subplot of Figure \ref{fig:eval_sds_mds_combo_IRMs} explains the drastic impact of the sound on offset detection, after which genre doesn't make much of a difference.
Similarly, 3rd subplot helps us dissect the $1:2$ ratio between the average influences of sound ($\sim$10\%) and genre ($\sim$20\%) in the velocity detection metric.
The 4th subplot then illustrates how these two note-level recognition performance dynamics compound. 
Ultimately, all three of them reveal how the T5 and Toyama systems are impacted disproportionately more strongly than their peers by the sound shift, in both offset and velocity detection.
The overall impact of genre-induced shift seems to be quite uniform among the systems.

Figure \ref{fig:eval_sds_mds_combo_MIMs} shows the average performance of each model on the two audio renditions of MAEtest, evaluated using musically informed piano transcription metrics. These metrics, in addition to the holistic perspective provided by the note-level information retrieval metrics, offer insight into how accurately specific musical dimensions are transcribed.

For articulation, which can be roughly understood as note duration, all models perform better on the original MAESTRO audio, regardless of whether the melody or bass stream is evaluated. The sound-related performance differences are slightly more pronounced in the melody stream, suggesting that changes in acoustic environment (e.g., timbral and overtone variations) might cause models to misinterpret higher-pitched notes more readily than lower-pitched ones. The effect of musical distribution shift seems to be minor, though this may be due in part to the simplicity of the genre-based sound samples compared to the more complex classical pieces in MAEtest.

In terms of timing—measured via inter-onset intervals (IOIs) to capture expressive onset patterns—all models, with the exception of OaF and Edwards in the melody stream, perform better on MAESTRO than on Disklavier audio. However, the performance drop in both melody and accompaniment seems minor. The performance gap between the two musical distributions (Classical in MAEtest, and the different genres in our MDS corpus) is rather drastic for most models, particularly in the melody stream. A notably poor result is observed for the T5 model in the melody stream. Unlike traditional note-wise F1 scores, timing metrics compute the correlation between sequences of onset values, emphasizing overall temporal alignment and expressive timing trends rather than discrete note-level accuracy. The result of T5 may stem from its token-based encoding scheme and the possibility of accumulated time shift errors.

Regarding the harmony-related metrics, sound distribution shift appears to have a mildly positive effect on most models -- especially Edwards -- whereas musical distribution shift has more of a negative impact across all models, which may reflect the models' greater sensitivity to changes in musical structure than to variations in timbral characteristics.
Finally, with respect to dynamics, all models demonstrate limited capability in capturing expressive loudness patterns independent of the type of distribution, with average correlation scores ranging from approximately 0.45 to 0.6. These results confirm that accurate modeling of dynamics or velocity remains an open challenge in piano transcription \cite{li2025piano}.

\subsection{Transcribing the Non-Musical: The Random Set}

To further expand our understanding of the music distribution shift and its effects on the rather extreme edge of the music distribution spectrum, we inspect how these AMT systems perform on our new collection of fully random musical sequences. Given the absence of meaningful musical structure in this data, we restrict our analysis to note-level metrics, offering insights into how corpus bias and entanglement problems surface in these far-OOD conditions.

The right half of Figure \ref{fig:eval_all_note_on_off_vel}
shows the average system performance on the triplets of pieces from the Random set, grouped by the degree of polyphony (number of concurrent voices) and laid out along the horizontal axis. 
Unsurprisingly, transcription performance degrades almost universally with growing number of concurrent voices, as there are simply more notes to catch, and thus also more to miss, especially given the lack of musical structure in the cross-talk of overlapping random voices.

\subsubsection{Note-level Performance Breakdown}
\label{sec:eval_note_rand}

\begin{figure}[t]
	\centering
	\includegraphics[width=\textwidth]{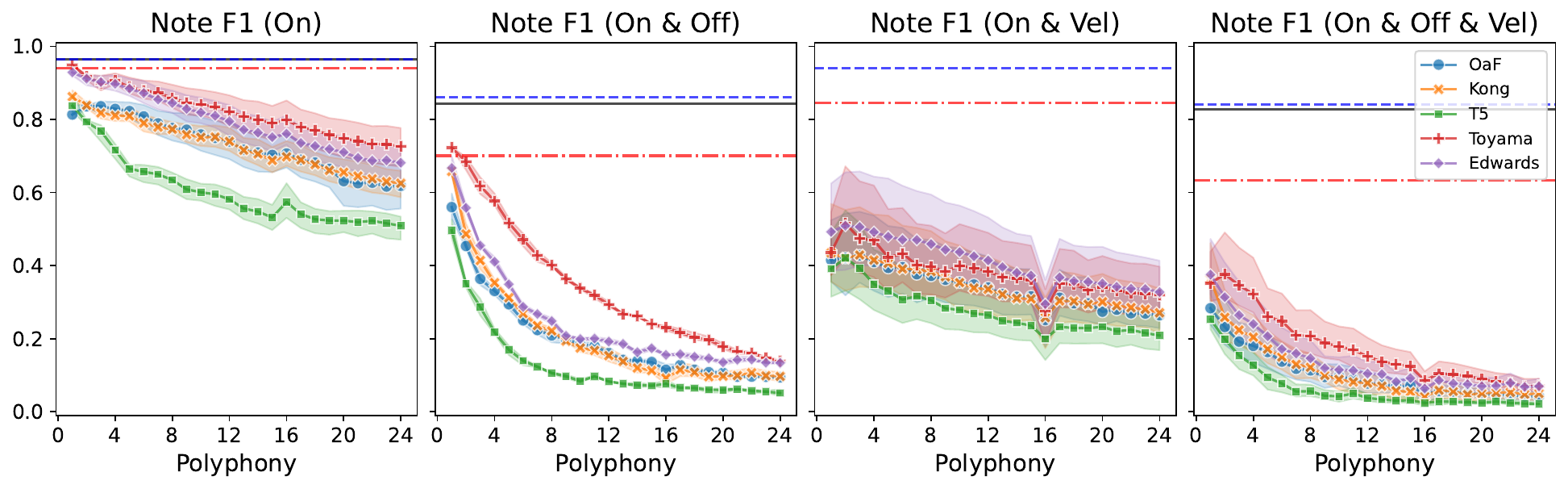}
	\caption{Performance of the evaluated AMT systems on the Random set of the MDS Dataset, broken down by the four variants of Note-level F1-score metric.}
	\label{fig:eval_random_IRMs_all}
\end{figure}

\begin{figure}[t]
	\centering
	\includegraphics[width=\textwidth]{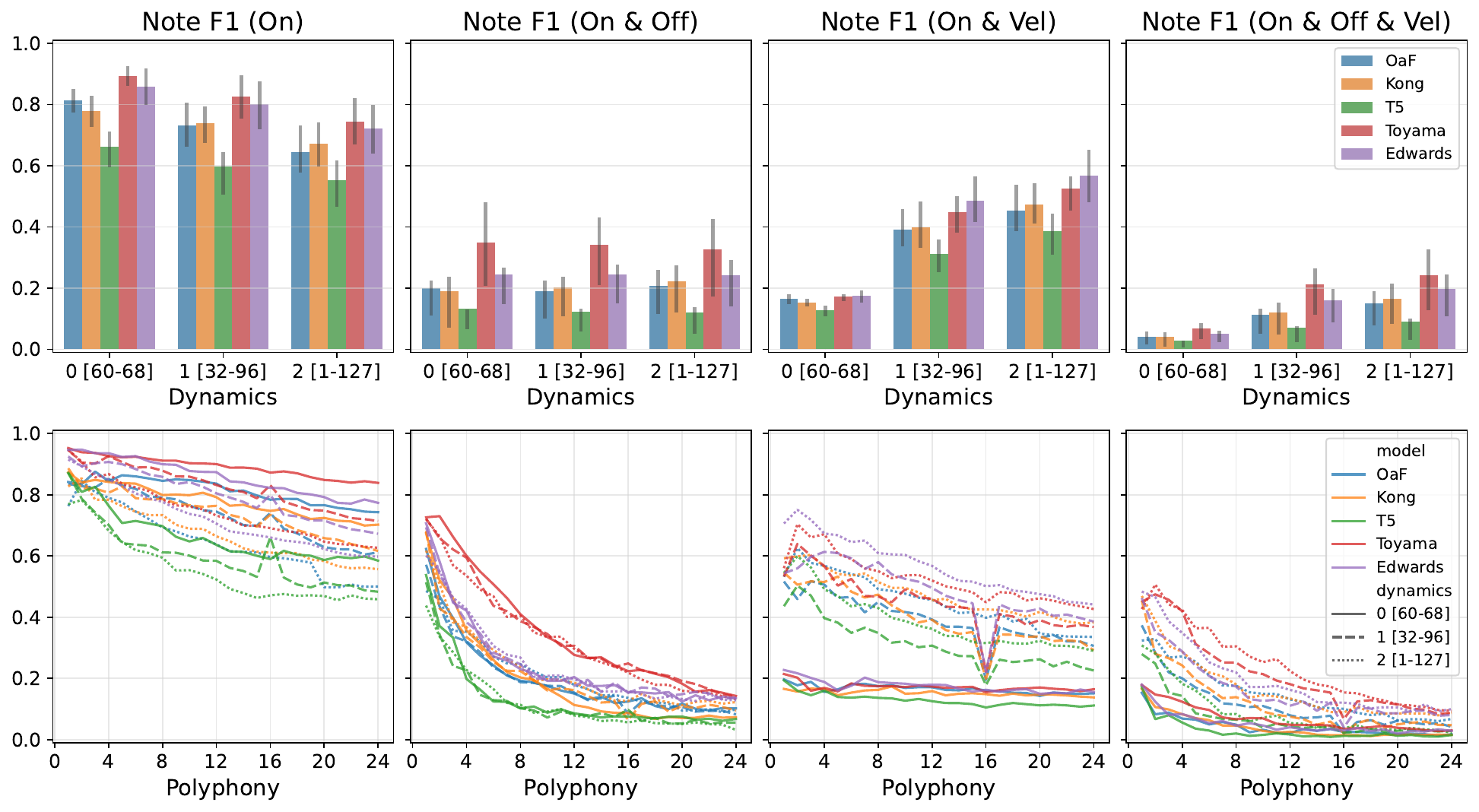}
	\caption{Transcription performance of the selected AMT systems on the (2) Random set. Columns show the Note-level metrics. First row shows bar plots aggregating results across the three levels of dynamics. Second row shows non-aggregated results, with line styles indicating dynamics.}
	\label{fig:eval_random_IRMs_dyn_breakdown}
\end{figure}

Figure \ref{fig:eval_random_IRMs_all} presents the Random set performance evaluation -- analogous to the right side of Figure \ref{fig:eval_all_note_on_off_vel}, but includes results for all Note-level F1-score metric variants, as described in Section \ref{sec:metrics_ir}.\footnote{The slight disturbance in performance trends (visible at polyphony level 16, and specific to the piece with ``medium'' dynamics -- level 1) in various metrics shown in Figures \ref{fig:eval_random_IRMs_all} and \ref{fig:eval_random_IRMs_dyn_breakdown} is likely due to a fresh tuning of our piano shortly before a (re-)rendering of this particular piece, which had become necessary due to an apparent recording failure in the main recording session.}
The first three subplots of Figure \ref{fig:eval_random_IRMs_all} can help us better understand how AMT systems handle the sub-tasks of complete note detection (last subplot).
The horizontal lines give us again reference for the average model performance on the MAESTRO data, following style and semantics of Figure \ref{fig:eval_all_note_on_off_vel}.
The rightmost subplot reproduces the data shown on the right half of Figure \ref{fig:eval_all_note_on_off_vel}, facilitating a side-by-side comparison.

\paragraph{On}
On the level of Onsets, we see a quasi-linear trend in most systems, as the data samples deviate further into the extreme.
While the Onset-only performance progressively degrades with growing polyphony in all 5 systems, one might have expected to see a bit more dramatic degradation, given the complete absence of musical structure in the underlying note sequences.

Perhaps, due to the highly pronounced spectro-temporal features of piano note onsets, that when temporally de-correlated -- as in our Random set -- they are more likely to be independently recognized by these systems.
Additionally, the absence of any musical structure in our Random pieces could be effectively ``bypassing'' any Music Language Modelling \cite{sigtia2016endtoend} that these systems are inherently learning, lowering the potential for any particular notes' surrounding spectro-temporal context to bias its individual estimation.

\paragraph{On \& Off}
With Offset detection (On \& Off) we see a rather quick deterioration of systems' performance, as the degree of polyphony grows.
This effect is also clearly non-linear, most of the degradation occurring within the first 8-12 polyphony levels.
Moreover, this metric reveals so far the clearest separation between the ``winning'' system -- Toyama, and the rest.

In terms of features, an Offset is manifested as a sudden halt of ongoing spectral activity.\footnote{Its ``suddenness'' being subject to possible reverberation, which can make precise Offset detection even more difficult. This observation is also specific to the piano, as different instruments may have ADSR templates with longer release phase.}
In contrast to Onset, this might be more easy to ``miss'' on an individual basis, especially for softly played notes.
However, if this was the sole cause of the degradation, one would expect to see different performance drops for varying rates of softly played notes (levels of dynamics); but our breakdown of its impact (see Figure \ref{fig:eval_random_IRMs_dyn_breakdown}) shows no such effect.
Additionally, while the analogous drop on the Genre set seems to be uniformly around $\sim$ 20\% (Figure \ref{fig:eval_genre_note_triplet_breakdown}), here we see a similar drop already at polyphony of 1, and only further strongly increasing from there.
This, along with the polyphony distribution in the Genre set concentrating around levels 3-5 (Figure \ref{fig:genre_polyphony_heatmap} in Section \ref{sec:data_genre}), suggests a likely presence of other factors contributing to these systems' otherwise high capacity for Offset detection.

One possible explanation is, that models trained on MAESTRO learn to rely on patterns found in \textit{musical} note sequences (herein absent) for estimating note offsets.
This would mean, that the aforementioned music language modelling might be detrimental for offset detection in such a far-OOD contexts as we have here, suggesting a link to the corpus bias problem.
Alternatively, the mere over-representation of longer notes in our Random set compared to the other musical subsets (upper-left quadrant of Figure \ref{fig:datastats_midi_all_juxtaposed}) could also partially explain a portion of the discrepancy between these two performance trends.

\paragraph{On \& Vel}
With the addition of Velocity detection (On \& Vel), in contrast to Offset, we see a consistent performance drop of $\sim$ 40\% across all polyphony levels, preserving the linear-like rate of deterioration.
This suggests, that -- unlike Offset -- the Velocity estimation is rather independent of polyphony.
Since the MAEtest results suggest that $\sim$ 10\% of the drop is sound-related, the remaining $\sim$ 30\% must be related to the (non-)musical material.
Similar to error bars in Figure \ref{fig:eval_genre_note_triplet_breakdown}, data bands here display increased performance variation among the levels of dynamics.
This is also the one metric, which flips the order of top two models on the Random set: Edwards outperforms Toyama (though by a small margin).

Figure \ref{fig:eval_random_IRMs_dyn_breakdown} additionally breaks these results down by level of dynamics.
We see a paradox, where predicting velocity levels in less dynamic pieces seems to be more difficult than in highly dynamic ones.
While Onset-only metric shows predictably lower performance on more dynamic content, and the added difficulty of Offset detection almost fully obscures this effect, it is completely reversed in the remaining two metrics that involve Velocity.
This could be a particular manifestation of the corpus bias problem, where systems trained on a particular distribution of MIDI note velocities then tend to map any loudness-related input features onto this expected distribution at test time.\footnote{As a relevant context, the bottom-left quadrant of Figure \ref{fig:datastats_midi_all_juxtaposed} depicts the stark contrast between the velocity distributions of ``musical'' material -- Genre and MAEtest -- and the narrow dynamics portion of our Random set (velocity range [60-68]).}
Indeed, our analysis of the velocity distributions produced by the AMT systems in Section \ref{sec:random_velo_dist} supports this interpretation.

\subsection{Noteworthy Observations}
We conclude this section with a few additional observations that extend beyond the main results presented above.

\subsubsection{Distributions of Note Velocity Values in the Random set}
\label{sec:random_velo_dist}

\begin{figure}[t]
	\centering
	\includegraphics[width=\textwidth]{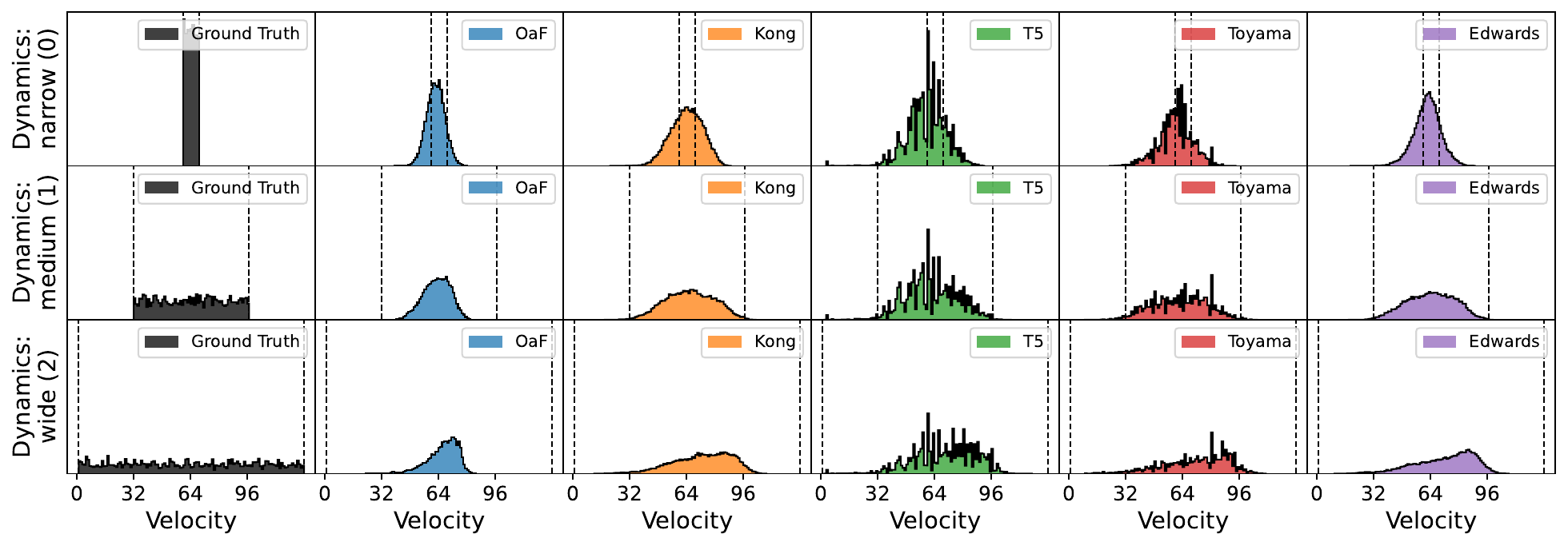}
	\caption{Histograms of note velocities from the Random set. Rows show subsets corresponding to the three levels of dynamics. First column shows ground truth data, followed by transcriptions of individual AMT systems. Vertical dashed lines designate the declared velocity ranges.}
	\label{fig:random_velocity_hist}
\end{figure}

We analyzed the transcriptions of all pieces in our Random set by counting velocities of notes, separately for each of the three levels of dynamics --- narrow (0), medium (1), and wide (2) --- and each of the five AMT systems.
The results are shown as Histograms in Figure \ref{fig:random_velocity_hist}, with the addition of ground truth in the first column, to highlight the differences between true and predicted distributions.

Data of the narrow (0) dynamics explain the counter-intuitive result from Figure \ref{fig:eval_random_IRMs_dyn_breakdown}, where systems generally show way worse performance in velocity-aware metrics on the narrow (0) dynamics subset, than the other two.
All five systems show a Gaussian-shaped statistic, reflecting their training data distribution.
This leads to a lot of predictions outside the range.
In the medium (1) dynamics, all systems' predictions are mostly within the true range.
Edwards, Toyama, and Kong seem to predict more evenly across the range.
The wide (2) range of dynamics shows a general lack of predictions at the edges of the range, which are -- as we know -- strongly underrepresented in the training data.
While systems differ in their limited capacities to predict in the full range, all systems also display some degree of skew to the right --- a tendency to predict louder notes more often.

Given that -- as Figure \ref{fig:eval_random_IRMs_dyn_breakdown} revealed -- performance at wide (2) dynamics still slightly improved from medium (1) in all systems, it is possible to imagine that a kind of ``anchoring'' effect takes place: where higher dynamic range in the input audio gives the models useful contextual reference, for what the edges of ``silent'' and ``loud'' sound like, resulting in enhanced precision.

Furthermore, the spikes in histograms of Toyama and T5 systems indicate that some velocity values are ``attractors'' -- inherently more likely to be predicted, than their neighbors -- which suggests a coarser velocity estimation capacity.
In contrast, the smooth histograms of OaF, Kong and Edwards suggest more even predictions of neighboring values, indicating a finer-grained estimation capacity in these systems.
Additionally, OaF seems to be the most constrained to predicting around the ``middle'' range (mezzo-forte).

\subsubsection{Relative Performance of the AMT Systems}
\label{sec:relative_model_perf}

\begin{figure}[t]
	\centering
	\includegraphics[width=\textwidth]{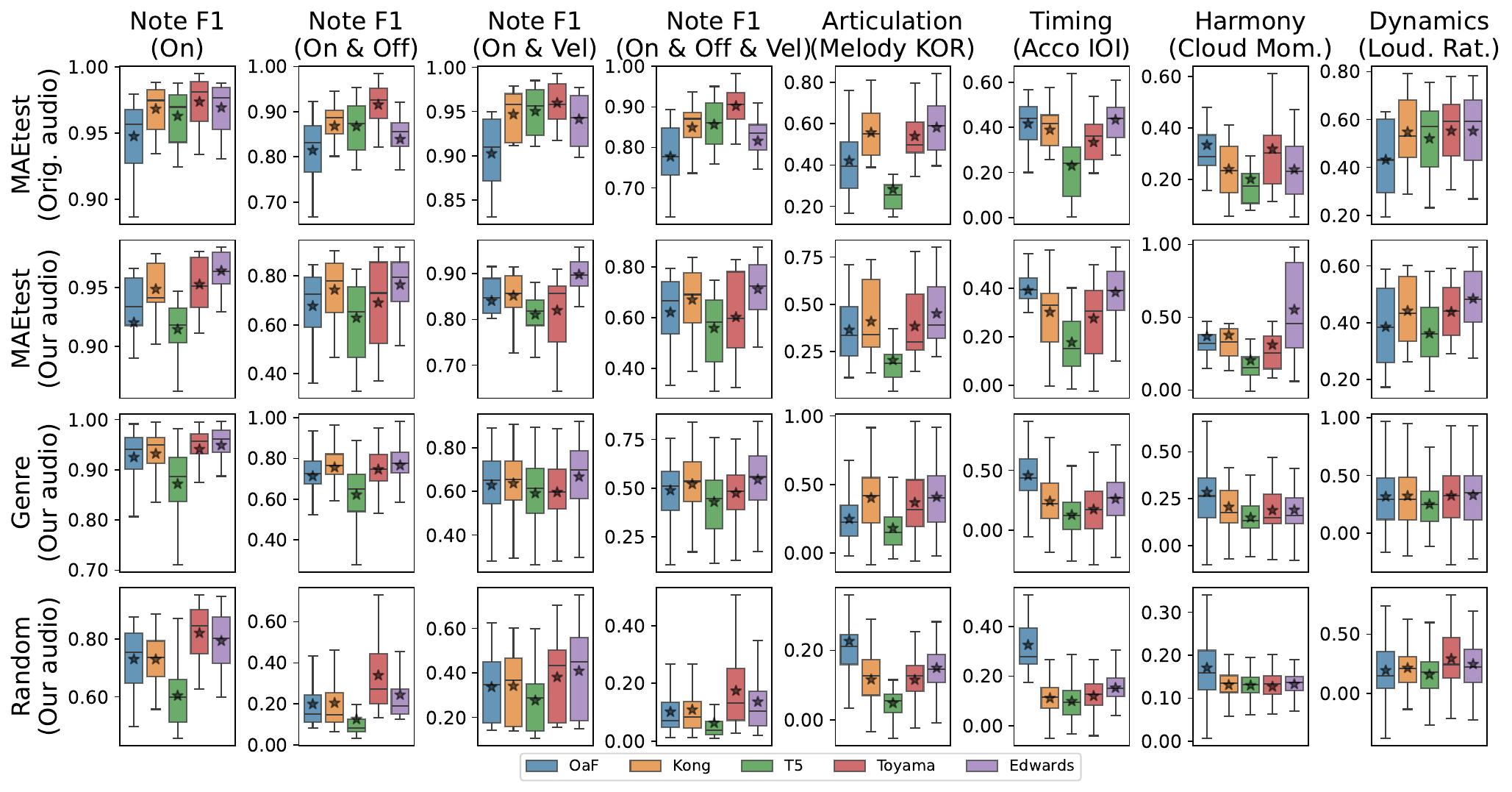}
	\caption{Relative performance of models across the 4 subsets of MDS Dataset (rows) in terms of 8 different metrics (columns): 4 types of note-level F1-score, and 4 selected musically informed metrics. Colors of the boxes denote AMT systems, following the color legend from Figure \ref{fig:inkscape_models_legend_aug}. The stars denote means, while the lines show medians.
	}
	\label{fig:eval_grid_data_metrics_rel_mod_perf}
\end{figure}

Figure \ref{fig:eval_grid_data_metrics_rel_mod_perf} displays how the five AMT systems perform on the aggregated subsets of the MDS Dataset (rows of the grid), in terms of different metrics (columns of the grid), relative to each other -- via visual proximity of candles.
Following the candles of specific systems within each column allows one to spot how relative performance in particular metric changes as the data distribution shifts (e.g. Toyama and Edwards in columns 1---4).
The sizes of candles illustrating the distribution of values within a metric can also help in setting expectations about performance consistency of the off-the-shelf systems in a particular context.
To illustrate a use case: a practitioner interested in note-level precision of their transcriptions might -- depending on their data and priorities (correct estimation of offset vs velocity) -- select Toyama or Edwards, respectively.
Another practitioner -- seeking to use an AMT system in a musicological analysis project (e.g. as part of a larger pipeline) -- might care more about the preservation of harmony and timing of the accompaniment parts, and thus gravitate towards the otherwise non-obvious choice of the less recently developed OaF system.

\subsubsection{Frame-level Performance}

\begin{figure}[t]
	\centering
	\includegraphics[width=\textwidth]{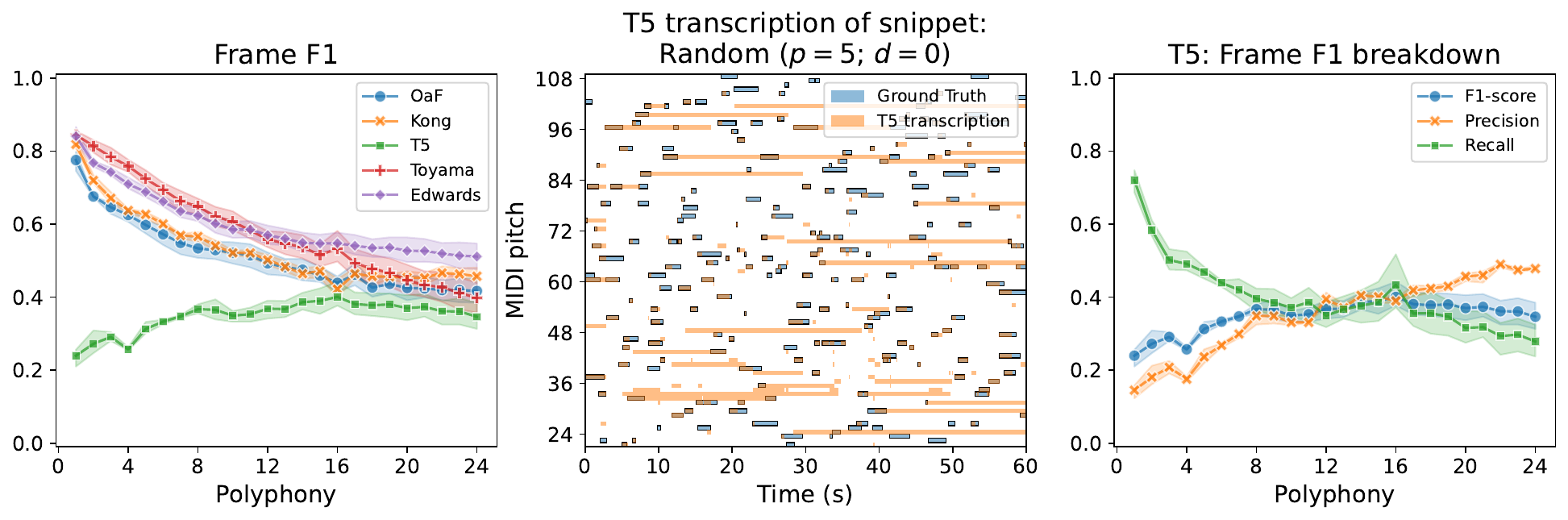}
	\caption{Results in Frame-level F1-score on the Random set (left) revealing an anomalous transcription performance of the T5 system (middle) further interpreted via precision/recall breakdown of the F1-score metric. The piano roll in the middle shows a 60 seconds snippet from the Random set (blue) overlaid with T5's transcription (orange): transparency defaults to 50\% instead of depicting velocity.}
	\label{fig:eval_random_frame_T5_anomaly}
\end{figure}

Since some of the AMT systems in our selection (and most modern AMT systems indeed) don't produce frame-level estimates, the frame-wise evaluation is becoming less relevant and informative\footnote{It quantifies a non-intuitive mixture of errors, resulting from: 1) inaccurate temporal placement of correct (true positive) notes, and 2) wrong (false positive) and omitted (false negative) ones.}.
Following common practice, we turn the note-level predictions into frame-wise representations\footnote{For this purpose, we compute piano-rolls from predicted MIDI at a temporal resolution of 100 frames per second.}.
The Frame-level F1-scores for the Random set are shown in Figure \ref{fig:eval_random_frame_T5_anomaly} on the left.
We see a rather anomalous performance curve for the T5 system: a very low start, followed by increase with growing with polyphony.

Upon inspection of the T5 transcription outputs (middle of Figure \ref{fig:eval_random_frame_T5_anomaly}), we found a clear explanation: due to the sequence-to-sequence nature of the T5 system and its tokenization of the (MIDI) output space -- note onsets and offsets are treated as separate, individual events in the vocabulary of tokens \cite{hawthorne2021sequence} -- anytime the system misses an offset, the note stays on until the next time an offset is eventually predicted for that note's pitch.
This occasionally leads to very long notes in T5's transcriptions, resulting in drastic increase in false positives on the frame-level, driving the precision measure down (Figure \ref{fig:eval_random_frame_T5_anomaly} right).
This effect seems to be inversely proportional to the polyphony level.
We conjecture, that this is due to growing density of notes, increasing chances of turning any particular forgotten note off, via its re-occurrence.

\section{Afterthought: DDS as an Alternative Solution to Corpus Bias}
\label{sec:dds}

The corpus bias problem was identified in \cite{martak2022balancing} when contrasting the performance of the dominant approach to AMT --  deep end-to-end supervised learning -- on the two subsets of the MAPS dataset \cite{emiya2010maps}: MUS (containing classical music pieces) and RAND (randomly generated chords).
In that work, we presented a method, named \textit{Differentiable 
	Dictionary Search (DDS)}, that was designed specifically to address the corpus bias problem.
The main idea was to combine the linear mixing model of \textit{Non-negative Matrix Factorization} \cite{lee1999learning}, which avoids the pitfall of memorizing training note combinations, with \textit{Normalizing Flows} \cite{tabak2010density} as highly flexible and expressive non-linear models of the instrument timbre and sound.
Due to limitations in scalability and computational costs, the system was never evolved to become competitive in note-level AMT with the deep end-to-end systems.

As an afterthought, we chose to additionally evaluate our DDS system -- the version presented in \cite{martak2022balancing} -- on our Random set, and put it in context with the results of AMT systems evaluated herein, to validate our original intuition that the DDS approach would be especially effective in situations of extreme music distribution shift.
Based on its rooting in nonnegative matrix factorization, the DDS system makes predictions at the frame level, as given by the magnitude spectrograms on its input.
To facilitate comparison at the note level, we extract notes after temporal smoothing of the predicted frame-wise activations, by using a simple 2-state Hidden Markov Model (HMM), similar to \cite{poliner2006discriminative}.
We use a fixed state transition model initialized as $(p_{\text{off}\rightarrow\text{on}}=0.30,\  p_{\text{on}\rightarrow\text{off}}=0.05)$.
The DDS activations are clipped to the range $[0; 1]$ and used as proxy for emission probabilities modelling the ``on''/``off'' states of our NoteHMM.
After Viterbi finds the most likely state sequence, we extract notes using the quantized time-stamps of frames with resolution of 32ms.

\begin{figure}[t]
	\centering
	\includegraphics[width=\textwidth]{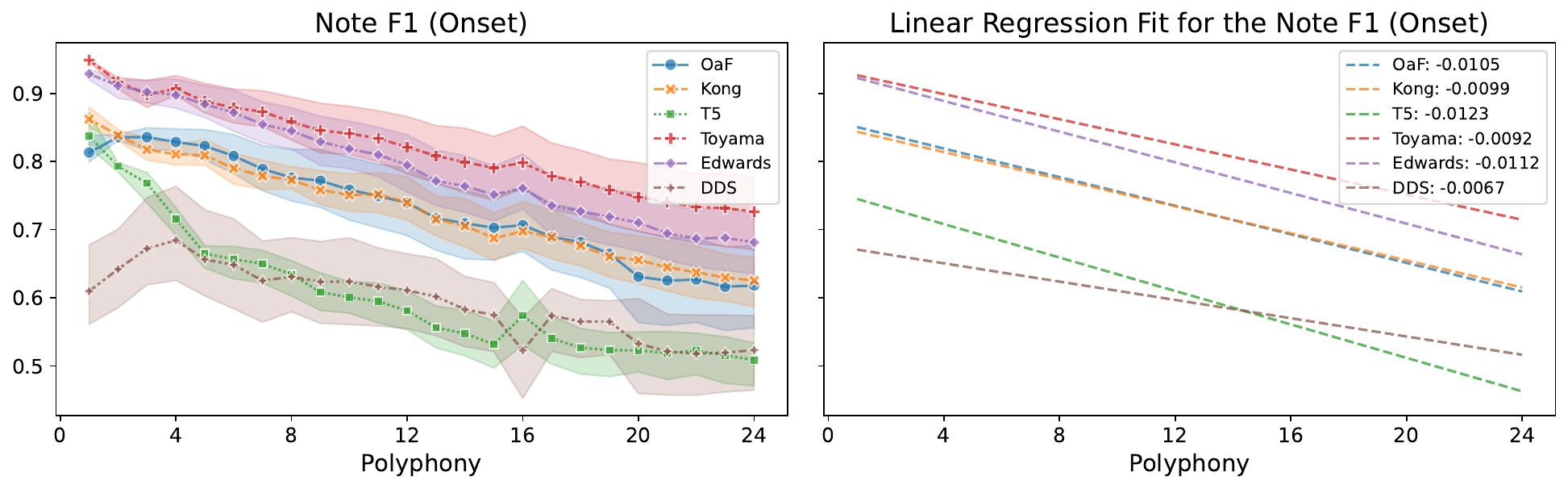}
	\caption{Note-level F1-score (Onset) evaluation of our DDS+HMM system on the Random set, in relation to DNN-based systems (left); linear regression fit to the performance curves along with their slopes in the legend (right).}
	\label{fig:dds_random_note_on_linreg}
\end{figure}

Since DDS decompositions extract information related to loudness and spectral energy as activations which are used exclusively as evidence for presence/absence of the notes in the NoteHMM, offset prediction is not competitive in this setup.
As a result, velocity would also need to be additionally estimated, which is beyond the scope of this experiment, as we are primarily interested in the dynamics of note entanglement and corpus bias.
The note-level evaluation in terms of the Onset-only metric is presented in Figure \ref{fig:dds_random_note_on_linreg}, where we juxtapose the DDS result against the suite of AMT systems evaluated throughout this paper.

We see an interesting performance growth until polyphony level 4, which contrasts the other systems.
Afterwards, performance slowly degrades in several steps, but appears to be kept slightly more stable overall compared to the DNN-based systems, for which performance keeps steadily decreasing as a function of polyphony.
The linear regression fit on these curves on the right side of the figure reveals that the DDS performance curve indeed has the slope with the smallest magnitude.
With all due caution, this can be interpreted as a sign of robustness to the extreme music distribution shift that manifests itself through increasing randomness of note combinations as a function of the polyphony (the number of randomly generated parallel voices) in our Random dataset.
Overall, however, DDS in its current form is not a practical alternative to current state-of-the-art AMT systems.

\section{Conclusions and Future Work}\label{sec:conclude}

In this paper, we presented an extensive evaluation of leading piano transcription models on out-of-distribution (OOD) data. To this end, we curated the MDS corpus, comprising three distinct subsets — (1) Genre, (2) Random, and (3) MAEtest — designed to assess inference performance under musical and sound-related distribution shifts, respectively.
Our analysis of musical distribution shift reveals strong evidence of corpus bias in both structured (genre) and unstructured (random) sequences, using standard note-level IR metrics and musically informed evaluations to uncover contributing factors. Specifically, we observe an average note F1-score performance drop of about 20 percentage points (p.p.), attributable to changes in sound, and an additional decrease of roughly 14 p.p. caused by shifts towards different genres, with deterioration reaching up to 50 p.p. on completely random sequences. Further breakdown of the Note F1 metric revealed surprising robustness of onset-only detection, while illuminating how additional recognition of offset time and note velocity remains a challenge in OOD settings. Specifically, offset detection seems to be much more impacted by changes in sound --- decrease of 16 p.p.  --- while velocity detection appears to be the most brittle under musical distribution shift --- decrease of 22 p.p. due to varying genres, and up to 50 p.p. when confronted with non-musical material.

An inherent limitation of any investigation of piano sound and reproduction concerns the sustain pedal calibration, which varies subtly between instruments. These variations arise from individual instrument configurations and pianist preferences, affecting how pedal pressure translates into mechanical action. As no universal ground truth exists for sustain pedaling behavior, this poses a challenge for consistent capture and reproduction, and should be considered in future dataset collection efforts.

Furthermore, we want to emphasize that genre classification is inherently fuzzy due to the broad and overlapping nature of genre boundaries, as well as the diversity within genre categories. This fuzziness, while complicating musical analysis, is important to accurately reflect the reality of musical style and practice. Therefore, we encourage the community to reflect this nuance in the development of future community benchmark datasets that systematically explore both musical and acoustic dimensions. Establishing such resources as community benchmarks would enable more robust evaluations of model generalization and help further disentangle the effects of musical versus acoustic variation.

\section*{Declarations}

\subsection*{Funding}
This work is supported by the European Research Council (ERC) under the EU’s Horizon 2020 research and innovation programme, grant agreement No. 101019375 ("Whither Music?"), by the LIT AI Lab, and by Johannes Kepler University Open Access Publishing Fund and the Federal State of Upper Austria.

\subsection*{Competing interests}
The authors declare that they have no competing interests.

\subsection*{Ethics approval and consent to participate}
Not applicable.

\subsection*{Consent for publication}
Not applicable.

\subsection*{Data availability}
The MDS dataset is made publicly available as a Zenodo repository at \url{https://zenodo.org/uploads/17467279}.

\subsection*{Materials availability}
Not applicable.

\subsection*{Code availability}
The source code used to generate and analyze the results for this study is made publicly available as a GitHub repository at \url{https://github.com/CPJKU/musical_distribution_shift}.

\subsection*{Author contribution}
LM designed the study and curated the dataset, PH administered the recording sessions and wrote the evaluation code. LM conducted the transcription experiments and refined the data, LM and PH jointly analyzed the results. GW supervised the study throughout. LM, PH and GW jointly wrote the manuscript. All authors read and approved the final version of this manuscript.

\subsection*{Acknowledgments}
We thank the authors of the five AMT systems evaluated here, for supplying the source code and model parameters of their transcription models.

\bibliographystyle{unsrt}
\bibliography{refs}

\newpage
\appendix
\section{Dissecting the MDS Dataset}
\label{sec:data}

As we wish to publish the MDS dataset and propose it as a new evaluation benchmark for future piano transcription research, it is important to understand and clarify its characteristics.
To that end, this appendix offers a detailed look into statistical/musical properties of MDS and its constituent subsets. %

\subsection{The Genre set}\label{sec:data_genre}
The Genre set contains 100 pieces which altogether contain 118,388 notes, with an average piece length of 3.14 minutes. %
We analyze this dataset by examining note and pitch class distributions, note density, interval patterns, and chord structures.

\paragraph{Pitch and pitch class distribution}

\begin{figure}[h]
	\centering
	\includegraphics[width=\textwidth]{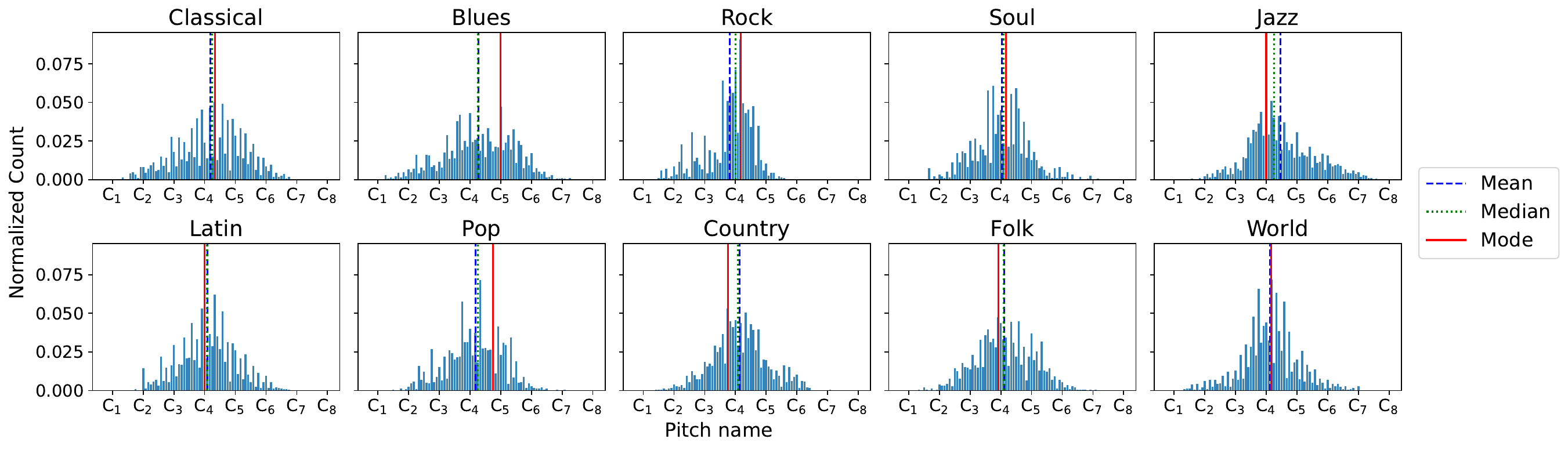}
	\caption{Note distribution per genre.}
	\label{fig:genre_note_distribution}
\end{figure}

First, Figure~\ref{fig:genre_note_distribution} presents the ten genres and their normalized note histograms. The horizontal axis represents the scientific pitch notation for the 88 keys of a modern piano, spanning from A0 to C8 in octave steps. The histograms demonstrate that the pitch distributions across all genres are approximately normal, with the majority of notes being centrally distributed. Notably, for most genres, except Rock and World, the predominant notes are concentrated within the octave range from C3 to C6. Cosine similarity analysis of the note distributions reveals most genres are highly similar, yielding a mean similarity score of $\mu$ = 0.875.

\begin{figure}[h]
	\centering
	\includegraphics[width=\textwidth]{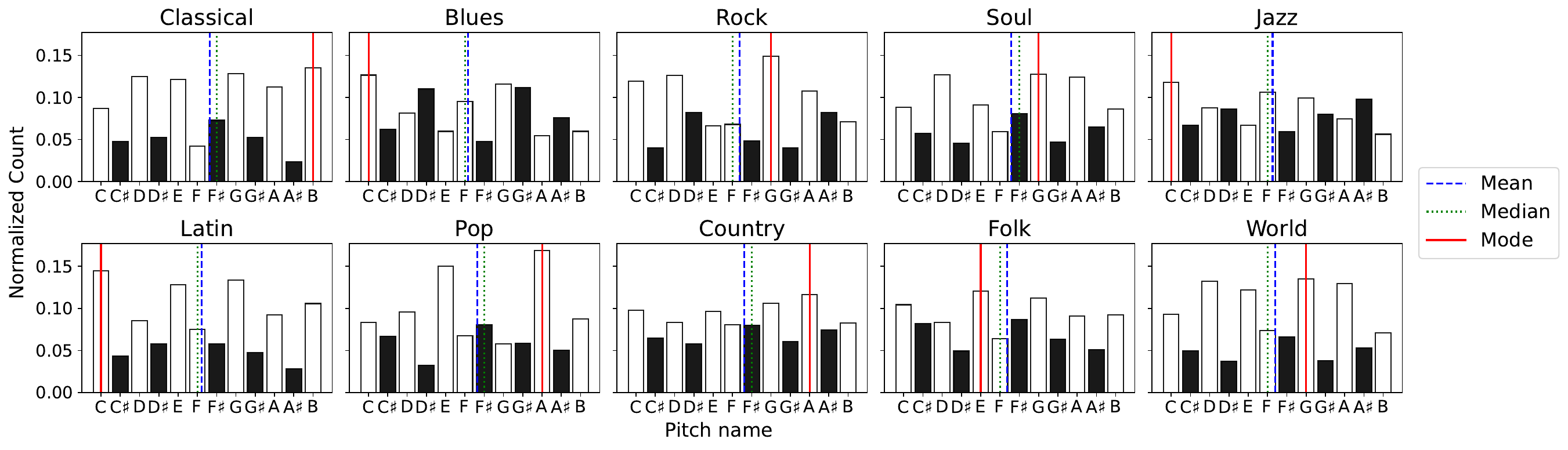}
	\caption{Pitch class distribution per genre.}
	\label{fig:genre_pitch_distribution}
\end{figure}

Figure~\ref{fig:genre_pitch_distribution} plots the normalized pitch class distributions across the ten genres studied, again with the horizontal axis representing scientific pitch class notation. Most genres center around pitch class $F$, and interestingly all genres show more frequent use of the natural ("white") pitch classes than the accidental ("black") ones, though this trend is more pronounced for the Classical, Pop, Latin and World Genre compared to Blues, Jazz and Country which tend to use the natural and accidental pitch classes more uniformly.

\paragraph{Density and polyphony levels}

\begin{figure}[h]
	\centering
	\includegraphics[width=0.7\textwidth]{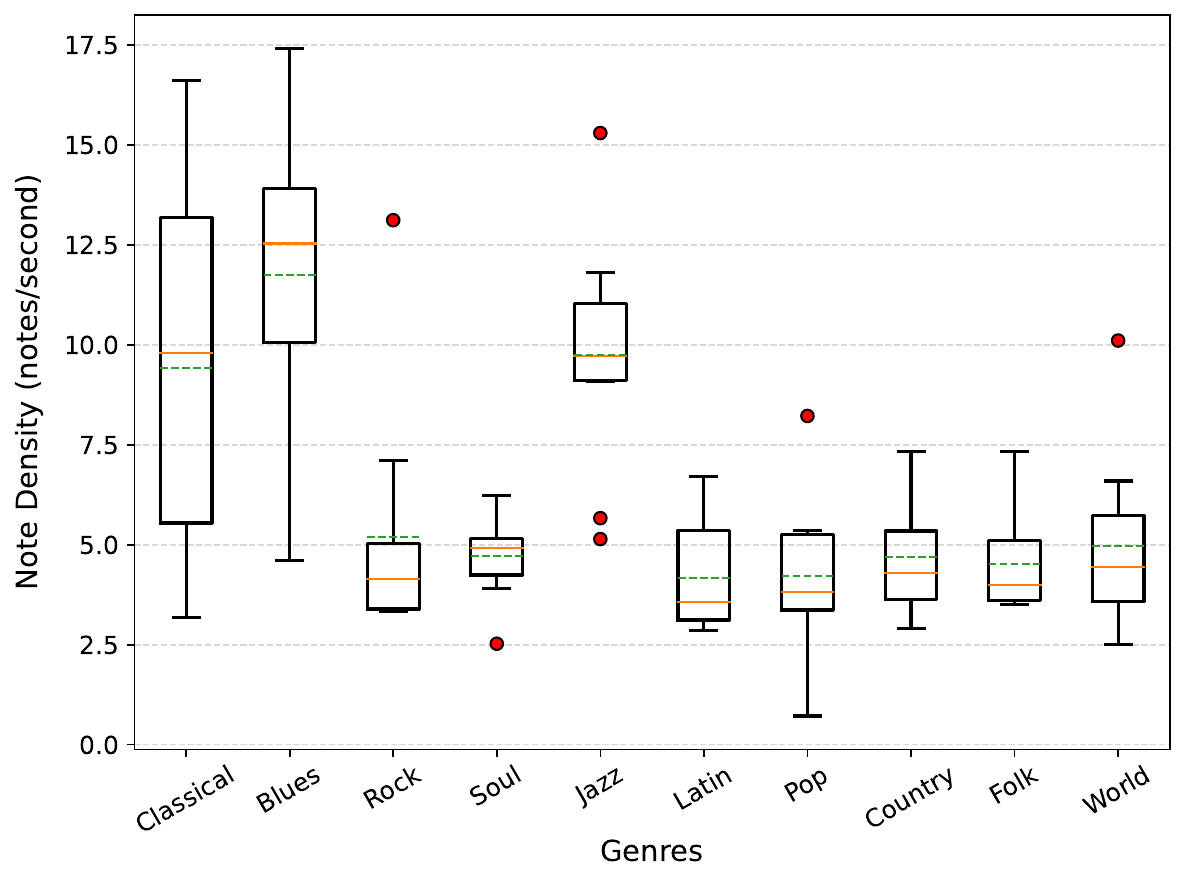}
	\caption{Box plot showing note density per genre.}
	\label{fig:genre_note_density}
\end{figure}

Next, Figure~\ref{fig:genre_note_density} presents the note density distribution (measured in notes per second) across genres, with each boxplot summarizing the variability within a genre.
Genres such as Blues and Classical exhibit a relatively wide interquartile range (IQR), indicating greater variability in note density compared to the remaining genres which show more tightly clustered distributions. Median note densities (indicated by the orange lines) vary significantly between genres; Blues, Classical and Jazz have higher medians, while the remaining genres exhibit lower central tendencies. Outliers are evident in genres such as Jazz, Pop, and World, suggesting occasional tracks with atypically high or low note densities. Overall, the Classical and Blues genres demonstrate both the highest densities and the broadest range. In contrast, genres like Folk and Soul are more homogeneous, e.g., characterized by consistently lower densities and small ranges, spanning primarily between 3–5 notes per second. This analysis highlights distinct patterns in note density across genres, reflecting inherent differences in musical complexity and structure.

\begin{figure}[h]
	\centering
	\includegraphics[width=\textwidth]{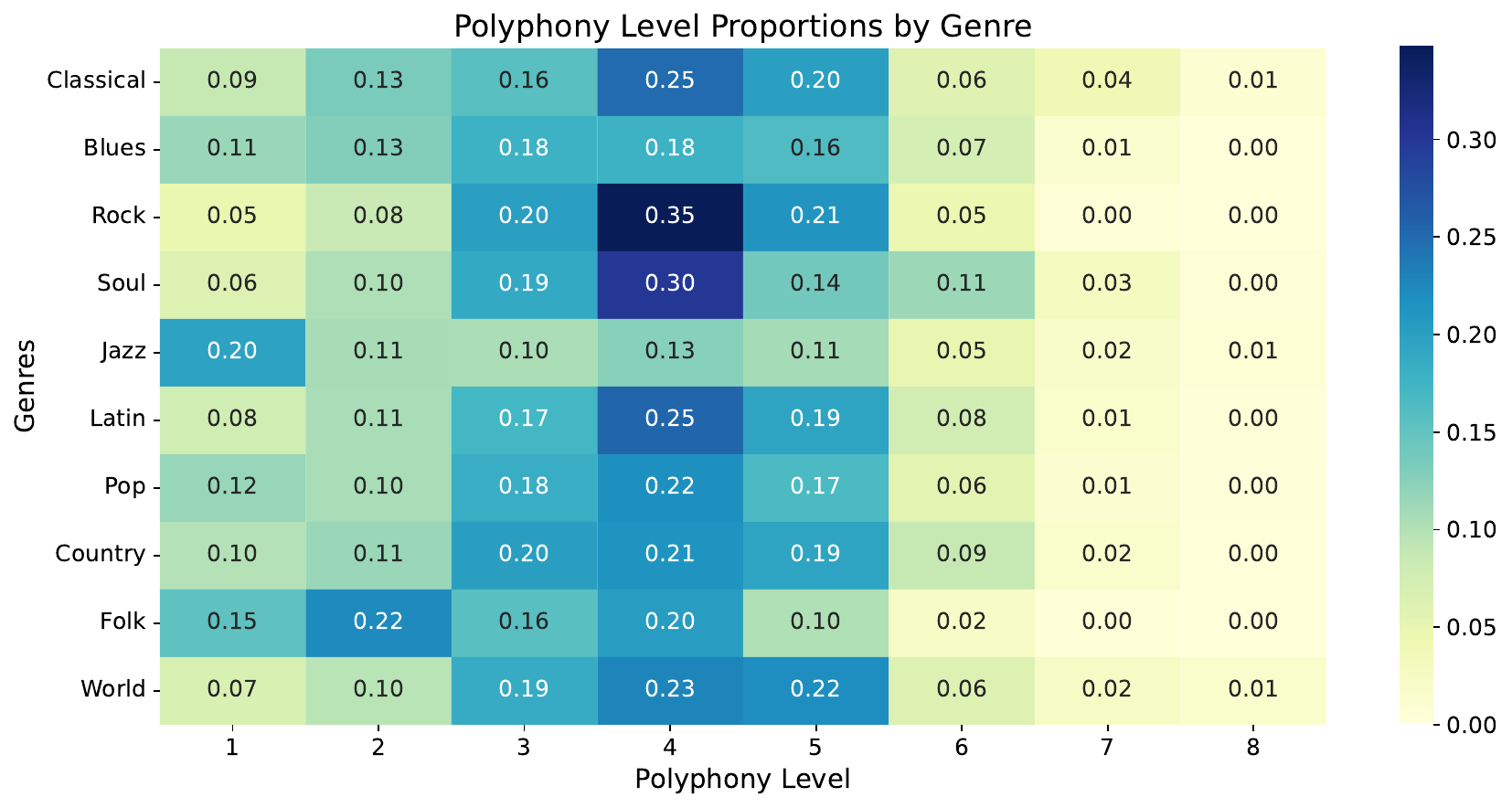}
	\caption{Normalized polyphony level count averaged within genre.}
	\label{fig:genre_polyphony_heatmap}
\end{figure}

Figure~\ref{fig:genre_polyphony_heatmap} illustrates the proportions of polyphony levels across ten different musical genres. The polyphony level represents the normalized count of simultaneous notes played at any given time, averaged within each genre. The horizontal axis denotes the polyphony levels, ranging from 1 to 8, while the vertical axis lists the genres. The heatmap reveals distinct patterns in polyphony usage among the genres. Jazz and Folk exhibit higher proportions of low polyphony levels ($p < 3$, effectively homophonic), while most other genres show higher proportions of mid-range polyphony levels (4-6), suggesting a higher frequency of more simultaneous notes. The highest polyphony levels (7-8) are less common across all genres, reflecting the rarity of extremely dense musical textures.

\paragraph{Intervals and chords}

\begin{figure}[h]
	\centering
	\includegraphics[width=\textwidth]{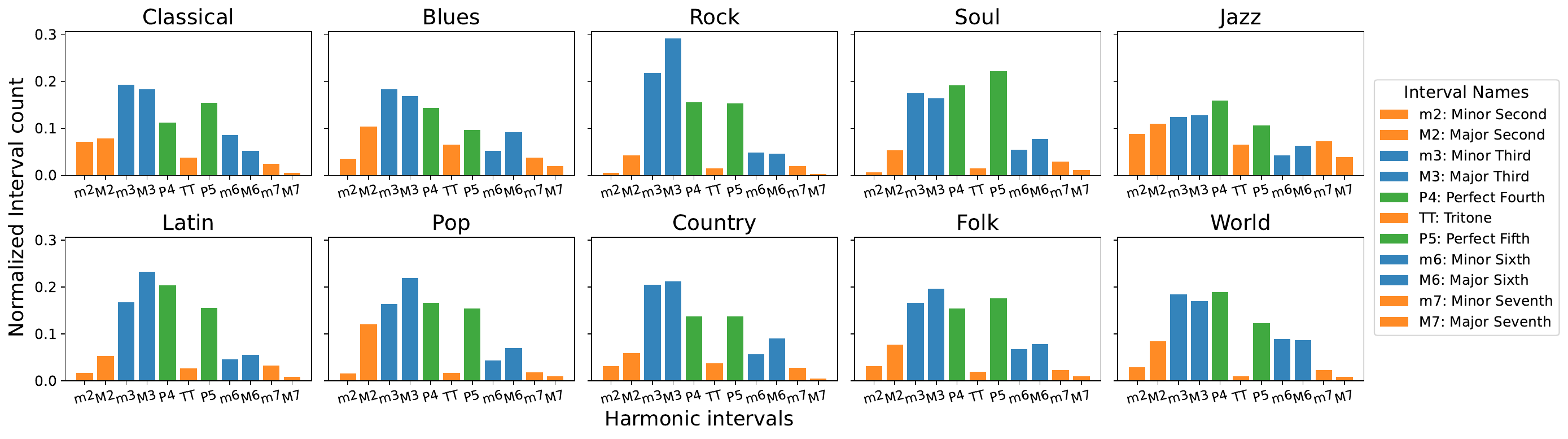}
	\caption{\textbf{Normalized interval count per genre}. The horizontol axis represents the harmonic interval, labeled using their common abbreviations: minor second (m2), major second (M2), minor third (m3), major third (M3), perfect fourth (P4), tritone (TT), perfect fifth (P5), minor sixth (m6), major sixth (M6), minor seventh (m7) and major seventh (M7). The vertical axis indicates the relative frequency of each interval in the genres studied. To enhance visual comparability, consonant intervals (P4, P5, P8) are highlighted in green, dissonant intervals (m2, M2, TT, m7, M7) are marked in orange, and all other intervals are shaded in blue.}
	\label{fig:genre_intervals}
\end{figure}

In the next part we analyse the interval and chord distribution across various genres. First, we are interested in the harmonic interval distribution. To this end, we count the occurrences of each unique ordered interval within the range of an octave. Only note combinations that occur quasi-simultaneously (within 20 milliseconds) are considered. Additionally, intervals larger than an octave are reduced to their equivalent form within the octave, after accounting for transpositional equivalence. 

Figure~\ref{fig:genre_intervals} illustrates the normalized harmonic interval counts across the ten musical genres considered. Across all genres, consonant intervals are generally more frequent than dissonant ones, highlighting their universal importance for establishing harmonic stability. This imbalance is less pronounced in the Jazz and related Blues genres, which exhibit a higher proportion particularly of tritones and seventh intervals, reflecting their emphasis on chromaticism and harmonic tension. Among the imperfect consonances, thirds are consistently more prevalent than their inverses (sixths), which is likely due to their harmonic importance in defining the quality of a chord in most tonal music. Notably, in many genres, thirds even surpass the perfect consonances in frequency, emphasizing their fundamental harmonic significance.

\begin{figure}[h]
	\centering
	\includegraphics[width=.8\textwidth]{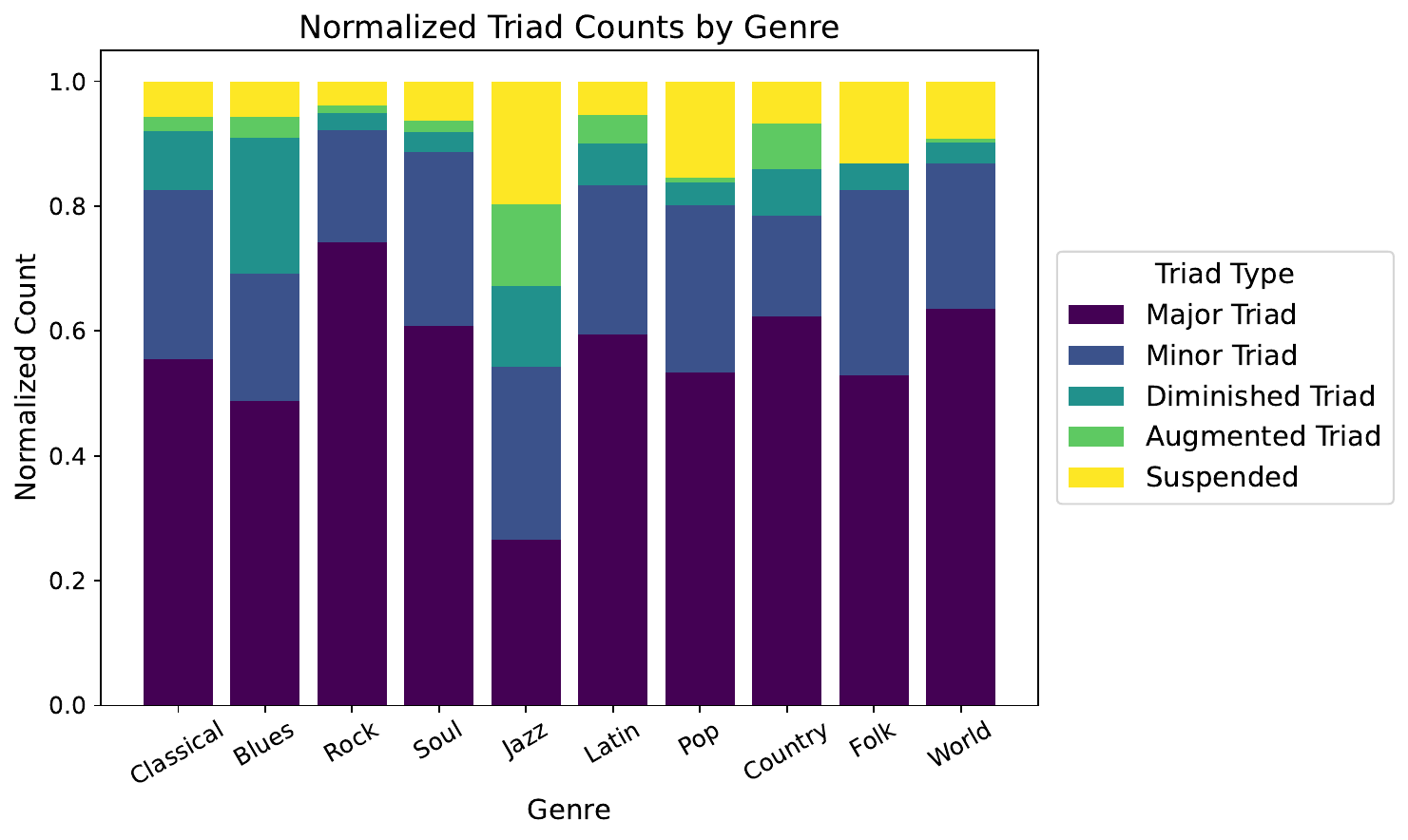}
	\caption{Normalized triad count per genre.}
	\label{fig:genre_triads}
\end{figure}

Chords, defined as the simultaneous combination of three or more notes, serve as fundamental harmonic structures in tonal and post-tonal music. They establish harmonic context within a given key or modality and can be categorized based on their constituent intervals. Traditional harmonic analysis often distinguishes between triads (three-note chords, including major, minor, diminished, and augmented forms) and seventh chords (four-note structures, or tetrachords, that introduce an additional seventh interval). 
Across different musical genres, chordal structures vary in complexity, frequency, and function, shaping the harmonic context characteristic of each style.

Figure~\ref{fig:genre_triads} illustrates the normalized counts of the most common triad types across ten different musical genres. The triad types include Major Triad, Minor Triad, Diminished Triad, Augmented Triad, and Suspended Triad, each represented by a different color. It can be seen that major triads are the most prevalent across all genres, followed by minor triads. Across all genres except Jazz and the related Blues genre, diminished, augmented, and suspended triads are the least frequent.

\begin{figure}[h!]
	\centering
	\includegraphics[width=.8\textwidth]{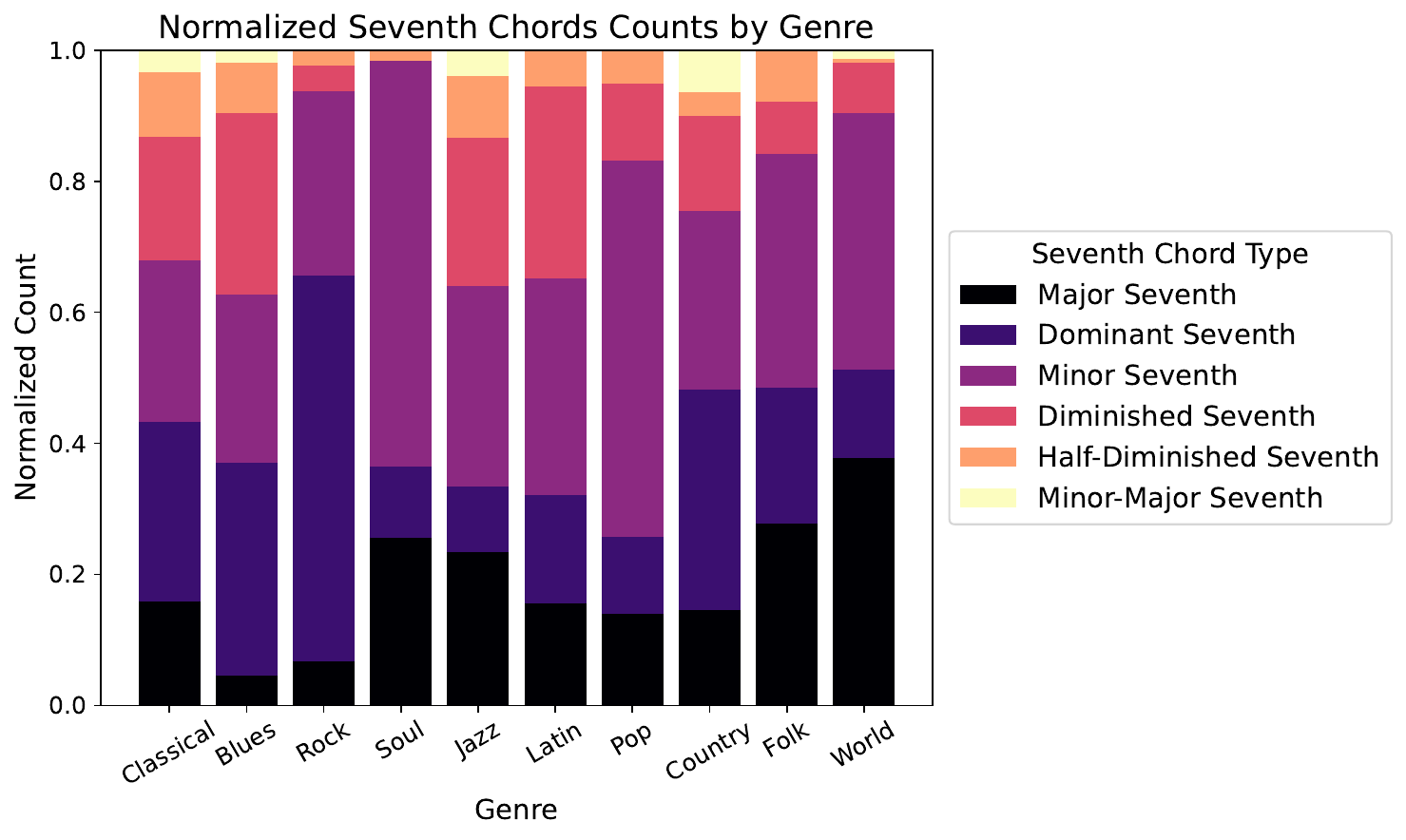}
	\caption{Normalized seventh chord count per genre.}
	\label{fig:genre_seventh}
\end{figure}

Figure~\ref{fig:genre_seventh} presents the normalized distribution of seventh chord types across genres. Each bar represents a genre, with stacked segments indicating the relative proportions of Major Seventh, Dominant Seventh, Minor Seventh, Diminished Seventh, Half-Diminished Seventh, and Minor-Major Seventh chords. 
Surprisingly, the minor seventh chord occurs most frequently across most genres, followed by the dominant, and the diminished seventh chord. The major seventh chord is relatively rare in most genres, except for the Folk and World genre, where it is more prevalent. The half-diminished and minor-major (minor third and major seventh interval) seventh chords are the least common types across all genres, with the exception of Jazz, where they are more prevalent.

\paragraph{Velocity}

\begin{figure}[h]
	\centering
	\includegraphics[width=\textwidth]{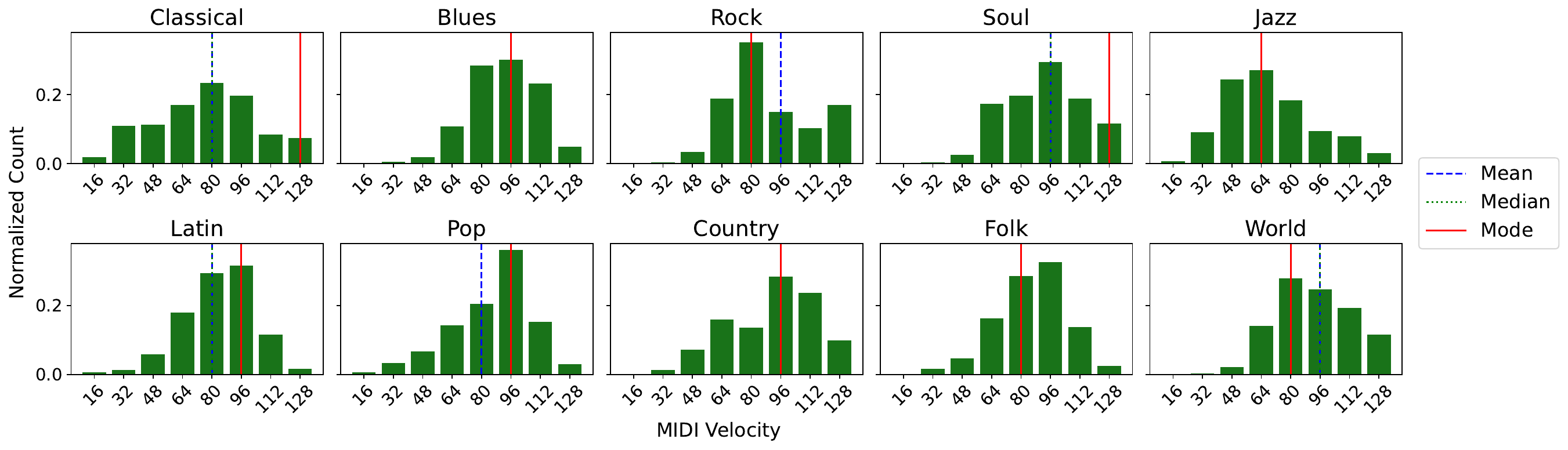}
	\caption{MIDI velocity histogram across genres.}
	\label{fig:genre_vel}
\end{figure}

Figure~\ref{fig:genre_vel} illustrates the distribution of MIDI velocity values across the ten genres, normalized for comparability. In each subplot, MIDI velocity values are binned, with the horizontal axis ticks describing the upper bound of each bin, and the y-axis representing their normalized count. The distributions are further summarized using the mean (blue dashed line), the median (black dotted line), and the mode (red solid line).

Most genres exhibit a slightly right-skewed distribution, with a peak in the mid-to-high velocity range (approximately 60–100). However, variations exist in the precise central tendency and spread. Notably, genres such as Classical and Jazz exhibit a wider range of velocities, indicating a more dynamic performance style with both soft and loud notes. Other genres, such as World, Blues, Rock, and Pop, show a more compressed velocity range, suggesting a more uniform dynamic level.

It should be noted that the plot represents MIDI velocities only, which serve as a proxy for loudness but do not directly reflect the actual produced or perceived volume. Factors such as instrument timbre, velocity sensitivity, and audio mixing can significantly influence the relationship between MIDI velocity and perceived dynamics, meaning that interpretations should consider these limitations.

\subsection{The Random set}
\label{sec:data_random}

\begin{figure}[h]
	\centering
	\includegraphics[width=\textwidth]{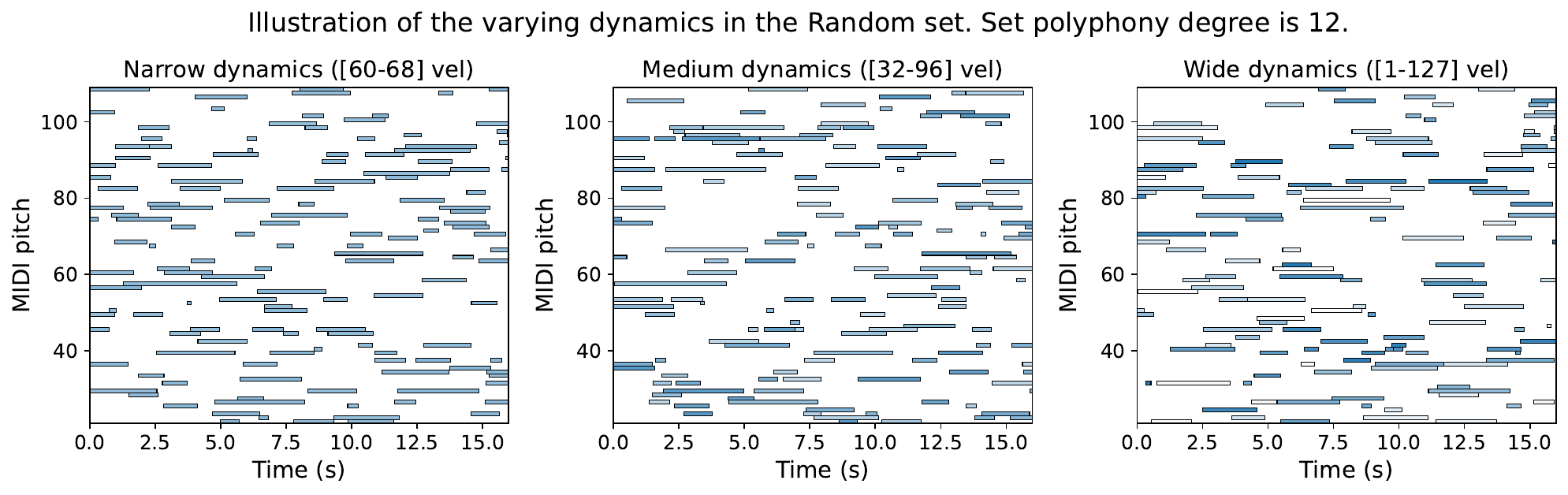}
	\caption{Illustrative samples of randomized note sequences for the Random set. Note velocities are depicted as transparency of the notes -- higher loudness means darker color (lower transparency).}
	\label{fig:random_dynamics_showcase}
\end{figure}

As the Random set is generated procedurally, its statistics are pre-determined and thus not so interesting for a musicological analysis.
Each piece with its given polyphony level and a level of dynamics is produced as a uniquely randomized set of parallel note streams.
This means that the three note sequences in a triplet of equal polyphony, but different dynamics, are not the same note sequence with different velocity levels of notes, but three unique note sequences.
To illustrate the content of the MIDI data underlying the Random set, Figure \ref{fig:random_dynamics_showcase} shows the piano rolls for a triplet of short snippets sampled with the same polyphony level of 12, and the three different ranges of possible dynamics.

\begin{figure}[h]
	\centering
	\includegraphics[width=\textwidth]{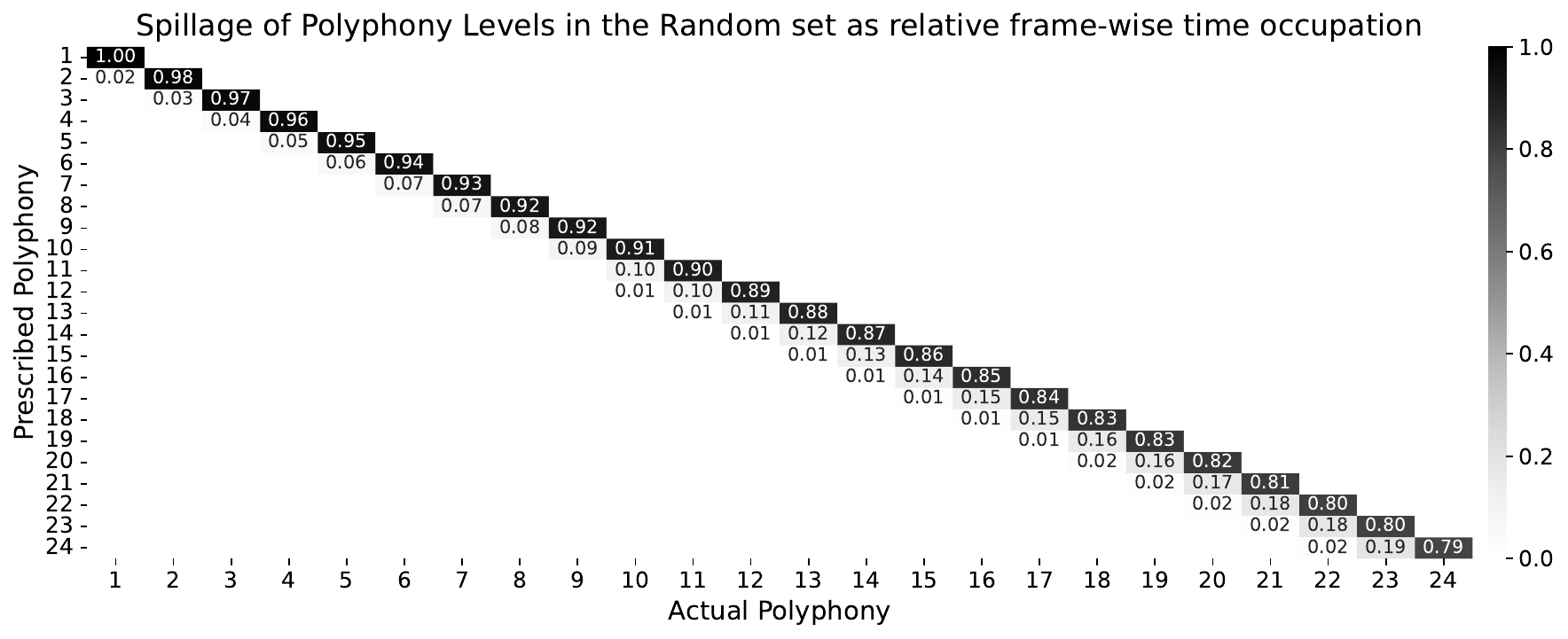}
	\caption{The effect of voice discontinuities on the maintained polyphony in our Random set.}
	\label{fig:random_polyphony_spillage_heatmap}
\end{figure}

As per the side effect of the Random sets' voice discontinuities introducing momentary drops in polyphony, mentioned in Section \ref{sec:curating}, we additionally visualize the negligible scope of this effect -- the ``spillage'' of prescribed polyphony level downwards, as the gaps in growing number of polyphonic voices compound.
Figure \ref{fig:random_polyphony_spillage_heatmap} documents this effect for the randomized trimming of up to 2\% per note, which was picked as a trade-off.

\end{document}